\documentclass[%
 aip,
 jmp,%
 amsmath,amssymb,
 reprint,%
]{revtex4}

\usepackage{graphicx}
\usepackage{dcolumn}
\usepackage{bm}
\newtheorem{theorem}{Theorem}
\newtheorem{corollary}{Corollary}

\newtheorem{lemma}{Lemma}
\newtheorem{definition}{Definition}

\newtheorem{remark}{Remark}

\newcommand{\comment}[1]{}

\newcommand{\qedwhite}{\hfill \ensuremath{\Box}}

\usepackage{graphicx}
\usepackage{amssymb}
\usepackage{amsmath}
\usepackage{eufrak}
\usepackage{float}
\usepackage{hyperref}
\usepackage{wasysym}
\usepackage{marvosym}
\usepackage{ifsym}
 \usepackage{stmaryrd} 
\usepackage{xcolor}

\begin{document}
%
\title{ Message Transmission over Classical Quantum
Channels     with a Jammer with Side Information:
  Correlation as Resource,  Common Randomness Generation
}
\author{Holger Boche}
\affiliation{ Lehrstuhl f\"ur Theoretische
Informationstechnik, Technische Universit\"at M\"unchen,
Munich, Germany\\
Munich Center for Quantum Science and Technology (MCQST), Munich, 
Germany
}
\email{boche@tum.de}
\author{Minglai Cai}
\affiliation{ Lehrstuhl f\"ur Theoretische
Informationstechnik, Technische Universit\"at M\"unchen,
Munich, Germany}
\email{minglai.cai@tum.de}
\author{Ning Cai}
\affiliation{ School of Information Science 
 and Technology, 
ShanghaiTech University, 
Shanghai, China} 
\email{ningcai@shanghaitech.edu.cn}

\begin{abstract}
In this paper we
analyze the capacity of a general model for arbitrarily varying classical-quantum
  channels (AVCQCs) when the sender and the
receiver use  correlation as a resource.
In this general model, a jammer  
has side information about the channel input.
    We  determine  a single letter formula
for the correlation assisted capacity.

As an application of our main result, we determine the
correlation assisted 
common randomness generation capacity.
     In this scenario the two channel users have access to
 correlation as a resource, and 
further use an  AVCQC with an
informed jammer for
additional discussion. The
goal is to create common randomness
between the two channel users.
     We also analyze 
    these capacity formulas     when
only  a small number of signals from the correlation are available.

For the correlation assisted 
common randomness   generation   capacity,
we show an additional interesting
property: For a sufficient amount
of ``public communication'',
common randomness generation capacity
is Turing computable, however
without this public communication
constraint, 
 the correlation assisted 
common randomness generation capacity
is, in general, not  Turing computable.
     Furthermore, we show that even without knowing the capacity formula of
the deterministic
capacity  using maximal error criterion, we can show 
that it is impossible to evaluate the performance algorithmically
on any current or future digital computer.
  
\end{abstract}

\maketitle


%


\section{Introduction}

Quantum information theory is a new field that allows us to exploit 
new possibilities, while at the same time impose fundamental 
limitations. 
In this work we consider 
classical-quantum channels.
 The capacity
of classical-quantum channels has been  determined  in   
\cite{Ho},   \cite{Sch/Ni},
and  \cite{Sch/Wes}.

The arbitrarily varying channel  (AVC) was  introduced
 in \cite{Bl/Br/Th2}.
This model describes 
  communication 
 including a jammer who tries to disturb the channel users' 
 communication  by changing his input in every channel use (cf. Figure \ref{CmAwtjhnxI}).
This model completely captures all possible jamming attacks which
depend  on the knowledge of the jammer.
     In \cite{Ahl/Bli} and \cite{Bj/Bo/Ja/No},
the capacity of arbitrarily varying classical-quantum channels is analyzed.

It is understood that the sender and the receiver have to select
their coding scheme first, and that this coding scheme is known by the
jammer, who chooses the most advantaged jamming attacking strategy
depending on this knowledge. In the conventional model in all above-mentioned works,
it is assumed that 
the jammer has no knowledge about the codeword which the sender sends.

Resources shared by users play an important role in the coding theorems for the AVC 
(cf. Figure \ref{awrarwtjhn}).
    For example, in  wireless communication, the communication
service may send some signals via satellite to its users.
In 1978 Ahlswede demonstrated in \cite{Ahl1}  the importance of 
the resources (of shared randomness) in a very clear form by
showing the surprising  result
  that either the deterministic capacity of an
arbitrarily varying channel is zero, or it is equal to its randomness assisted
 capacity (Ahlswede dichotomy).


In \cite{Bo/No}, a classification of various resources is given. A distinction is made between  two extremal cases:
randomness and correlation.    Randomness is the strongest resource, it
requires a perfect copy of the
outcome of a random experiment, and thus we should assume
an additional perfect channel to generate this kind of  resource.
On the other hand,  the weakest resource is defined as follows.
Assume that a bipartite source, modeled by an i.i.d. 
(independent and identically distributed) random variable $(V',V)$  
with values in a finite product set ${\mathbf V}'\times{\mathbf V}$, is observed by the 
sender and  receiver. The sender has access to the random variable $V'$ and 
the receiver to $V$. We call $(V',V)$ a correlation.

The work \cite{Bo/No} showed
 that the common randomness is a stronger resource than the correlation in the following sense:
a sufficiently large amount of common randomness  allows the sender and receiver to 
asymptotically simulate the statistics of any correlation. To the contrary,  
an example is given where not even a finite amount of common 
randomness can be extracted from a given  correlation without further communication.

As already mentioned, in the above-mentioned works it is assumed that the
jammer knows the coding scheme, but has no side
information about the codeword.
In many applications,
especially for the secure communications, it is too optimistic
to assume this.
In \cite{Bo/Ca/Ca} it has been shown that the jammer can
benefit from his knowledge about the sent codeword,
i.e, he may have a better jamming strategy.
Thus in  our
previous paper \cite{Bo/Ca/Ca} we
 considered the scenario
when the jammer  knows both the coding scheme and the input
codeword.

This work is an   extension   of our
previous paper
\cite{Bo/Ca/Ca}, where we 
determined the randomness assisted capacity of 
  AVCQCs with
a jammer
knowing the channel input (cf. Figure \ref{awtjkhract}).
 However, as mentioned above,  common randomness  is  a very ``costly'' resource.
 A
promising result of this work is that 
 the much
``cheaper'' resource, the
  correlation, is also an equally powerful
    resource (cf. Figure \ref{awtjntcshm}). 
 Furthermore, a correlation
$(V',V)$ does not have to be ``very good'' to be helpful in achieving a positive
secrecy capacity, since $(V',V)$ is a helpful resource even if
$I(V',V)$  is only slightly larger than
zero.
We also show that the same
capacity can been achieved 		using
a smaller amount (as compared to the number
of channel uses) of correlation.

As an application of our results, we turn to the
question: How much common randomness can an
  AVCQC with
an informed jammer  
generate, using the correlation as a resource?
     In this scenario, the two  channels users 
have, as a resource, access to a correlation which is
characterized by a random variable $(V',V)$,  
with values in a finite product set ${\mathbf V}'\times{\mathbf V}$.
The sender has access to the output of $V'$ and 
the receiver to $V$. 
In addition,
the sender is allowed to send  helper messages
via an  AVCQC with an
informed jammer. The
goal is to create shared random variables 
 with negligibly small errors.  
The amount of shared random variable per channel use
is denoted by the common randomness generation capacity.
     Capacities of common randomness generation 
 over classical perfect channels and
over classical noisy channels have been
determined in \cite{Ahl/Cs}.
In this work, we deliver the
common randomness generation capacity 
with an informed jammer using correlation as the resource.
We also analyze 
the case when only a smaller amount (as compared to the number
of channel uses) of correlation is used.

In 1912, Borel attempted  to express   concepts of 
computability in \cite{Bo}. In \cite{Tu1} and  \cite{Tu2},
Turing introduced  the concept of computable
numbers and computable functions
 on computable real numbers. Computable numbers
are real numbers that are computable by a Turing machine, which
is a mathematical model of an abstract machine that
manipulates symbols on a strip of tape and  can simulate any given algorithm.
Problems which are not  computable by a Turing machine cannot be
solved or even algorithmically approximated by any current or future digital computer.
Banach and Mazur defined in \cite{Ba/Ma}
 a notion of computability for  functions on real numbers,
 using  computable sequences.

The Turing computability of a capacity
is a necessary condition for evaluating the performance of
the corresponding system. Otherwise, it is impossible to
algorithmically evaluate or even approximate it by computable
continuous functions.
\cite{Bo/Sch/Ba/Po} 
considered secret key capacities and  secure authentication
capacities over several classical channel network models, and 
determined whether they are computable, i.e.
if they can be algorithmically solved 
with the help of Turing machines. 
     As a further application of our results, we 
  now   analyze  some capacity formulas of this work
and determine whether they are Turing computable.
     We show that whether the
common randomness generation capacity 
with informed jammer is  Turing computable
depends on the amount of correlation
the channel users have access to. We also 
analyze  the Turing computability of a capacity whose
formula is still unknown.     
\vspace{0.3cm}

\section{Definitions and Communication Models}

\subsection{Basic Notations}

Throughout the paper, the random variables will be  denoted by capital
letters, e. g. $S,X,Y,$. Their realizations (or values) and
domains (or alphabets) will be denoted by the corresponding lower case letters,  e.g.
$s,x,y,$ and script letters, e.g. ${\cal S},{\cal X},{\cal Y}$, respectively. Random 
sequences will be denoted   by the capital bold-face letters, whose lengths are understood by the context, e.g.
${\bf S}=(S_1, S_2, \ldots, S_n)$ and ${\bf X}=(X_1,X_2, \ldots, X_n)$, and deterministic sequences 
will be written as lower case bold-face letters, 
e.g.  ${\bf s}=(s_1,s_2, \ldots, s_n), {\bf x}=(x_1, x_2, \ldots, x_n)$. 

$P_X$ is the distribution of the random variable $X$. 
Joint distributions and conditional distributions of random variables $X$ and
 $S$ will be written as $P_{SX}$, etc.  and $P_{S|X}$ etc., 
respectively, and $P_{XS}^n$ and $P_{S|X}^n$ will be  their product distributions, i.e.
$P_{XS}^n({\bf x},{\bf s}):= \prod_{t=1}^nP_{XS}(x_t,s_t)$, and 
$P_{S|X}^n({\bf s}|{\bf x}):=\prod_{t=1}^nP_{S|X}(s_t|x_t)$. 


Throughout the paper, dimensions of all Hilbert spaces are finite.
For a finite-dimensional
complex Hilbert space  ${\cal H}$, we denote
the (convex) set 
of  density operators on ${\cal H}$ by
\[\mathcal{S}({\cal H}):= \{\rho \in \mathcal{L}({\cal H}) :\rho  \text{ is Hermitian, } \rho \geq 0_{{\cal H}} \text{ , }  \mathrm{tr}(\rho) = 1 \}\text{ ,}\]
where $\mathcal{L}({\cal H})$ is the set  of linear  operators on ${\cal H}$, and $0_{{\cal H}}$ is the null
matrix on ${\cal H}$. Note that any operator in $\mathcal{S}({\cal H})$ is bounded. \vspace{0.3cm}

Throughout the paper, the logarithm base   is  $2$.      For 
a discrete random variable $X$  on a finite set ${\cal X}$ and a discrete
random variable  $Y$  on  a finite set   ${\cal Y}$,   we denote the Shannon entropy
of $X$ by
$H(X)=-\sum_{x \in \mathcal{X}}P_X(x)\log P_X(x)$ and the mutual information between $X$
and $Y$ by  
$I(X;Y) = \sum_{x \in \mathcal{X}}\sum_{y \in \mathcal{Y}}  P_{X,Y}(x,y) \log{ \left(\frac{P_{X,Y}(x,y)}{P_X(x)P_Y(y)} \right) }$.
Here $P_{X,Y}$ is the joint probability distribution function of $X$ and $Y$, and 
$P_X$ and $P_Y$ are the marginal probability distribution functions of $X$ and $Y$, respectively.



If the sender wants to transmit a classical message  set  to
the receiver using a quantum channel, his encoding procedure will
include a classical-to-quantum encoder 
to prepare a quantum state $\rho \in
\mathcal{S}({\cal H})$ suitable as an input for the channel. 
In view of this, we have the following
definition.\vspace{0.3cm}

\begin{definition}
Let $\mathcal{H}$ be a finite-dimensional
complex Hilbert space.
 A classical-quantum channel      is
a mapping ${\cal W}: \mathcal{X}\rightarrow\mathcal{S}({\cal H})$,
 specified by a set of quantum states $\{\rho(x), x \in {\cal X}\}$  $\subset\mathcal{S}({\cal H})$, 
indexed by ``input letters" $x$ in a finite set ${\cal X}$. ${\cal X}$ and ${\cal H}$ 
are called input alphabet and output space, respectively. 
We define the $n$-th extension of the 
 classical-quantum channel ${\cal W}$ as follows.
The channel outputs a quantum state
$\rho^{\otimes n}({\bf x}):=\rho(x_1) \otimes \rho(x_2) \otimes \ldots, \otimes \rho(x_n)$
in the $n$th tensor power ${\cal H}^{\otimes n}$ of the output space ${\cal H}$ when an 
input codeword ${\bf x}=(x_1,x_2, \ldots, x_n) \in {\cal X}^n$ of length $n$ is input into the channel.
\end{definition}
 \vspace{0.3cm}

For a quantum state $\rho\in \mathcal{S}(H)$ we denote the von Neumann
entropy of $\rho$ by \[S(\rho)=- \mathrm{tr}(\rho\log\rho)\text{
.}\]

Let
${\cal W}$: $\mathcal{X} \rightarrow
\mathcal{S}({\cal H})$ be a classical-quantum
channel.   For $P\in P(\mathcal{X})$,  
the conditional entropy of the channel for ${\cal W}$ with input distribution $P$
is presented by
 \[S({\cal W}|P) := \sum_{x\in {\cal X}} P(x)S({\cal W}(x))\text{
.}\]

Let $\Phi := \{\rho_x : x\in \mathcal{X}\}$ 
be a classical-quantum
channel, i.e. a
 set of quantum  states
labeled by elements of $\mathcal{X}$. For a probability distribution  $Q$
on $\mathcal{X}$, the    Holevo $\chi$ quantity is defined as
\[\chi(Q;\Phi):= S\left(\sum_{x\in \mathbf{A}} Q(x)\rho_x\right)-
\sum_{x\in \mathbf{A}} Q(x)S\left(\rho_x\right)\text{ .}\]

\subsection{Communication Scenarios}
\label{xodndijtucltaictakfn}

In this subsection we 
introduce our communication concept
and some resource models whose
capacity results (well-known from previous works) we
need for this work.

\begin{figure}[H]\begin{center}  
 \includegraphics[width=0.75\textwidth]{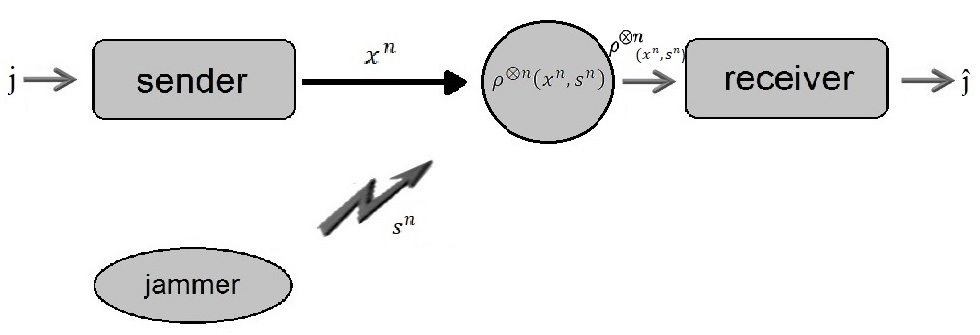}
 \caption{Conventional model: AVCQC when the jammer    has no
further knowledge about the channel input and the  channel users 
have no access to any resource: In this scenario
the jammer's inputs do not depend on $x^n$.}\label{CmAwtjhnxI}
\end{center}\end{figure}

\begin{definition}

An arbitrarily varying classical-quantum channel (AVCQC) ${\cal W}$ is specified by a 
set $\{\{\rho(x,s), x \in {\cal X}\}, s \in {\cal S}\}$ of classical-quantum channels with a common input alphabet ${\cal X}$ and output space ${\cal H}$ 
which are indexed by elements $s$ in a finite set ${\cal S}$. Elements $s \in {\cal S}$ are usually called the states of the channel.
${\cal W}$ outputs a quantum state
\begin{equation} \label{eq_f1a}
\rho^{\otimes n}({\bf x}, {\bf s}):=\rho(x_1, s_1) \otimes \rho(x_2, s_2) \otimes \ldots, \otimes \rho(x_n, s_n)
\end{equation}
if an input codeword ${\bf x}=(x_1,x_2, \ldots, x_n)$ is input into the channel 
and the channel is governed by a state sequence ${\bf s}=(s_1, s_2, \ldots, s_n)$,
 while the
state varies from symbol to symbol in an  arbitrary
manner.

\end{definition}

We assume that
the channel state $s$ is in control  of
the jammer.
Without loss of generality, we also assume
that the jammer always chooses the most advantageous attacking strategy,
     i.e. he wants the  channel users
	to transmit a least reliable message.     This is important for the applications
of our result to other channel models,
e.g. compound channels.
     The conventional AVCQC model when the jammer    has 
information about the channel input is shown in Figure \ref{CmAwtjhnxI}.
\vspace{0.2cm}

\begin{definition}\label{lmambfslxybdet}

An 
  $(n, J_n)$  code  $\mathcal{C}$ 
consists of an    encoder $u^n:
\{ 1,\cdots ,J_n\} \rightarrow {{\cal X}}^n$, and
 a set of collections of positive-semidefinite operators $\{D_j^n:
  j = 1,\cdots ,J_n\}$ on ${\cal H}^{\otimes n}$ which fulfills $\sum_{j=1}^{J_n} D_j^n =
\mathrm{id}_{{\cal H}^{\otimes n}}$.


\end{definition}
 \vspace{0.3cm}

\begin{definition}
A non-negative number $R$ is an achievable 
rate  for a classical-quantum channel
${\rho}(x)$ if for every $\epsilon>0$, $\delta>0$,
 and sufficiently large $n$ there exists an  $(n, J_n)$
code $\mathcal{C} = \bigl(u^n, \{D_j : j = 1,\cdots J_n\}\bigr)$,  such that $\frac{\log
J_n}{n}
> R-\delta$, and
\[1- \frac{1}{J_n} \sum_{j=1}^{J_n}
\mathrm{tr}\left({\rho}^{\otimes n}(u^n)D_j\right) < \epsilon\text{ .}\]

 The supremum on achievable deterministic 
   rates  of ${\rho}(x)$  is called the
  capacity of ${\rho}(x)$, denoted by
$C( {\rho})$.

\end{definition}

\begin{figure}[H]\begin{center}  
 \includegraphics[width=0.75\textwidth]{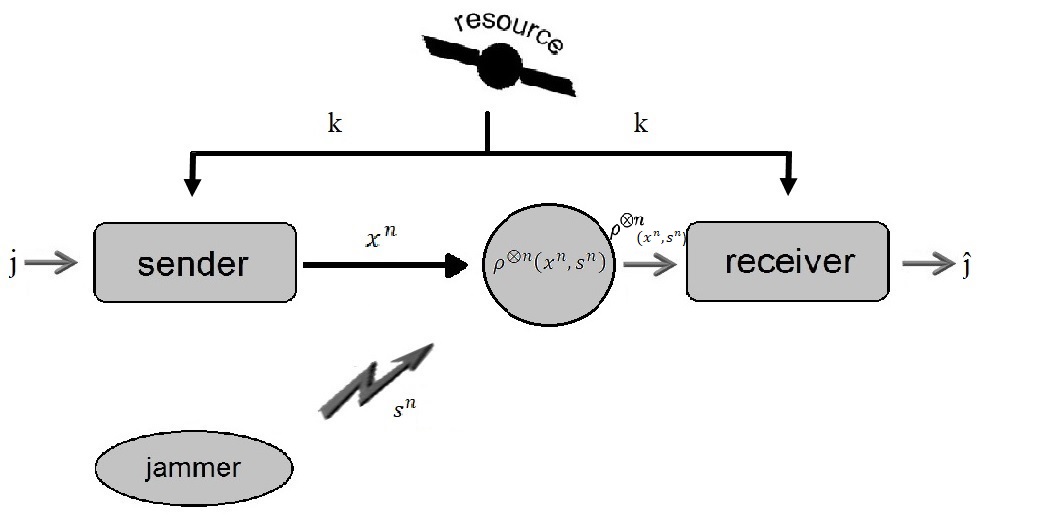}
 \caption{AVCQC  with randomness
as a coordination resource when the jammer    has no
further knowledge about the channel input.}\label{awrarwtjhn}
\end{center}\end{figure}

In the context of the arbitrarily varying channels, randomness can be an
important resource for a reliable communication over  an arbitrarily varying channel.   The
 message transmission task over AVCQC with common randomness
between the sender and the receiver is illustrated in Figure \ref{awrarwtjhn}.
   Ahlswede showed in \cite{Ahl1} (cf. also \cite{Ahl2} and  \cite{Ahl3}), the surprising  result
  that either the deterministic capacity of an
arbitrarily varying channel is zero, or it equals its randomness  assisted
 capacity (Ahlswede Dichotomy). 
\cite{Bo/Ca/De} shows there are indeed arbitrarily varying classical-quantum
 channels which have zero deterministic  capacity and positive random  capacity.
Therefore randomness  is indeed a very helpful resource
 for  message transmission (and secure message transmission) through an arbitrarily varying classical-quantum
 channel.
Having some resource is particularly essential for the scenario we consider
in this work (see the discussion below).

\begin{figure}[H]\begin{center}  
 \includegraphics[width=0.75\textwidth]{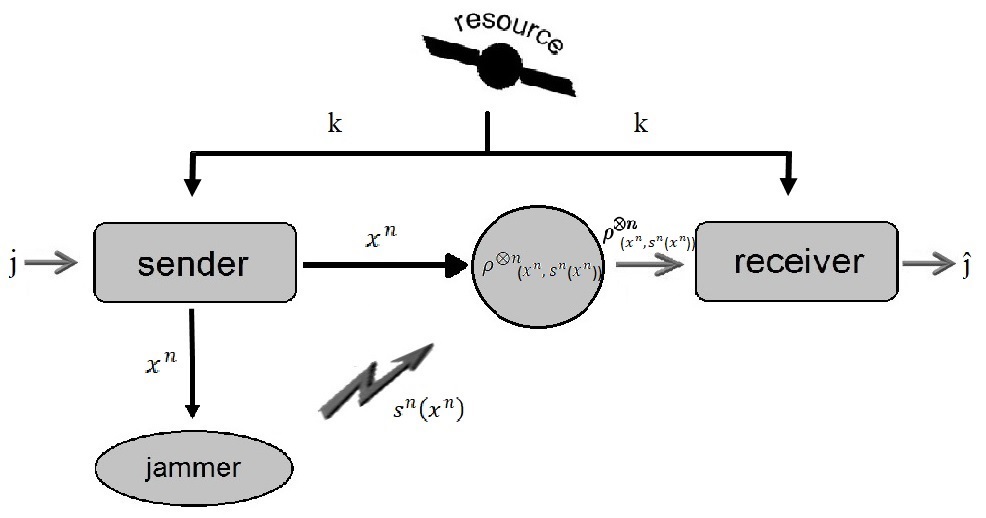}
 \caption{AVCQC  when the jammer knows the code word, while the
 users  have randomness
as a coordination resource: The sender and the receiver
share    the outcome of a random experiment, i.e. they share common randomness. }\label{awtjkhract}
\end{center}\end{figure}

Most of the previous works in AVCQCs
consider the case when the jammer knows the coding scheme,
but has no side information about the codeword of the  transmitters.
However,    \cite{Bo/Ca/Ca} shows that the jammer can strictly
reduce the capacity     when he knows the codeword. Thus
we concentrate on
message transmission over classical-quantum
channels with a jammer with additional  side information about the codeword.
We assume
that the jammer  chooses the most advantageous attacking strategy
according to 
his side information.
  The corresponding message transmission task is 
illustrated in Figure \ref{awtjkhract}.


 \vspace{0.3cm}
\begin{definition}
Let $\{\{\rho(x,s), x \in {\cal X}\}, s \in {\cal S}\}$ be an  AVCQC. 
 A non-negative number $R$ is an achievable  deterministic rate   
with an informed jammer (under the average error criterion) 
  for $\{\{\rho(x,s), x \in {\cal X}\}, s \in {\cal S}\}$, if for every
$\epsilon>0$, $\delta>0$,  and every sufficiently large $n$ there
exists a   code
$\mathcal{C} =
\biggl\{u^n,\{D_j^n: j\in\{
1,\cdots ,J_n\}\}\biggr\}$  such that $\frac{\log J_n}{n}
> R-\delta$, and
\[\max_{ {\bf s}^n (\cdot)}   P_e(\mathcal{C},  {\bf s}^n(\cdot)) < \epsilon\text{ ,}\]
where $P_e(\mathcal{C},  {\bf s}^n)$  is defined as 
\[ P_e(\mathcal{C},{\bf s}^n(\cdot)) := 1- \frac{1}{J_n} \sum_{j=1}^{J_n}
\mathrm{tr}(\rho(u^n(j), {\bf s}^n(u^n(j)))D_j^n)\text{ .}\] 
Here the maximum $\max_{ {\bf s}^n (\cdot) }$ is taken over all 
functions  ${\cal X}^n \rightarrow {\cal S}^n$.

 The supremum on achievable deterministic 
   rates  of
 $\{\{\rho(x,s), x \in {\cal X}\}, s \in {\cal S}\}$  with an informed jammer under the average error criterion is called the
deterministic    capacity of
 $\{\{\rho(x,s), x \in {\cal X}\}, s \in {\cal S}\}$  with an informed jammer, denoted by
$C( \{\{\rho(x,s), x \in {\cal X}\}, s \in {\cal S}\})$.

 \end{definition}

 Throughout the paper, 
all rates are defined,
unless otherwise stated, 
under the
average error criterion.

\vspace{0.3cm}

Our scenario (when the jammer knows the input codeword)
is already a 
challenging topic for classical arbitrarily varying channels.
This has been analyzed by Sarwate   in \cite{Sar}, where 
only the randomness assisted capacity has been determined.
The deterministic capacity formula, i.e. without additional resource,   
is even in the classical case an open problem.
It 
has been shown by Ahlswede in \cite{Ahl0}  that the classical  capacity 
under the maximal error criterion in this scenario  contains  the 
zero-error capacity of related discrete memoryless channels  
   as a special case.  A  deterministic capacity formula
for this is still unknown.
In particular, \cite{Bo/Ca/Ca} shows a violation of the Ahlswede dichotomy in  our scenario.

Coding for AVCQC  with an informed jammer is even
harder. 
Due to the non-commutativity of quantum operators, 
many
techniques, concepts and methods of classical information theory
may not be extended to quantum information theory. 
For instance, it is still unknown how to extend list decoding,
which has been used in \cite{Sar}'s proof, and non-standard decoder,
 to quantum information theory. 
In
\cite{Bo/Ca/Ca} we determined
the randomness assisted
capacities  of AVCQCs  when the jammer
has access to the channel input.
\vspace{0.3cm}

\begin{definition}

A randomness assisted  code $\Gamma$ for an AVCQC ${\cal W}$ is a uniformly distributed
random variable, taking values in a set of codes 
$\{({{\cal V}'}(k), \{{\cal D}(j,k), j \in {\cal J}\}), k \in {\cal K}\}$ with a common 
message set ${\cal J}$, where ${{\cal V}'}(k)=\{{\bf u}(j,k), j \in {\cal J}\}$ and $\{{\cal D}(j,k), j \in {\cal J}\}$ 
are the code book and decoding measurement of the $k$th code in the set, respectively.          $|{\cal K}|$
depends on the length of the codes in this set, i.e. it
 is a 
function of $n$. Particularly, for a fixed $n$,
  $|{\cal K}|$ is finite.          \label{arccg}
\end{definition}\vspace{0.3cm}

\begin{definition}

By assuming that the random message $J$ is uniformly distributed, we define 
\begin{align} 
&P_e(\Gamma)=\max_{{\bf s}}\mathbb{E}tr[\rho^{\otimes n}({\bf u}(J,K), {\bf s} ({\bf u}(J,K)))(\mathbb{I}_{\cal H}-{\cal D}(J,K))] \nonumber \\
&=\max_{{\bf s}}\frac{1}{|{\cal J}|} \sum_{j \in {\cal J}} \sum_{k \in {\cal K}} Pr\{K=k\}tr[\rho^{\otimes n}({\bf u}(j,k), {\bf s} ({\bf u}(j,k)))(\mathbb{I}_{\cal H}-{\cal D}(j,k))].
\end{align}
This can be also rewritten as
\begin{align}\label{eq_f2a} &
P_e(\Gamma)=\sum_{{\bf x}} Pr\{{\bf u}(J,K)={\bf x}\} \max_{{\bf s} \in {\cal S}^n}\mathbb{E}\{tr[\rho^{\otimes n}({\bf u}(J,K), {\bf s})(\mathbb{I}_{\cal H}-{\cal D}(J,K))]|{\bf u}(J,K)={\bf x}\}.
\end{align}

A non-negative number $R$ is an achievable  rate   for the
arbitrarily varying classical-quantum  channel
${\cal W}$  under randomness assisted   coding with an informed jammer, 
  if  for every
 $\delta>0$ and  $\epsilon>0$   and every sufficiently large $n$    there is a
 randomness assisted  code
 $\Gamma$  of length
$n$, such that
$\frac{\log |{\cal J}|}{n} > R-\delta$ and
$P_e(\Gamma) < \epsilon$.

 The
supremum on achievable rate under randomness assisted   coding   of ${\cal W}$ with an informed jammer
 is called the
randomness assisted   capacity of
 ${\cal W}$ with an informed jammer, denoted by
$C^{*}({\cal W})$.

\end{definition}\vspace{0.3cm}

\subsection{Code Concepts and Resources}

In this subsection we introduce the relevant code
concepts and resource models for this work.

\begin{figure}[H]\begin{center}  
 \includegraphics[width=0.75\textwidth]{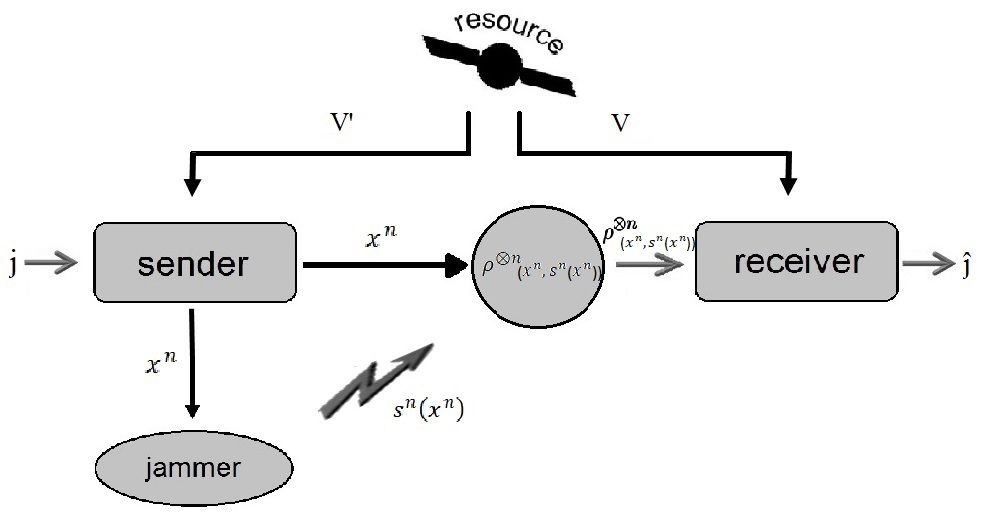}
 \caption{AVCQC  when the jammer knows the coding scheme, while the
 users   have only correlation
as a resource.}\label{awtjntcshm} 
\end{center}\end{figure}

A correlated source is a 
discrete memoryless source  (DMS) $(({V'}^n,V^n))_n$  observed by the sender and
receiver,  modeled by $n$ independent copies of a random variable
$(V',V)$, with values in some finite set ${\cal V}' \times  {\cal V}$. The sender has
access to the random variable $V'$, and the receiver to $V$.   The
corresponding message transmission task  over AVCQC with correlation
between the sender and the receiver is 
illustrated in Figure \ref{awtjntcshm}.
   We call 
$\Bigl(({V'}^n,V^n)\Bigr)_n$ a correlated source, or
a correlation. Since $\Bigl(({V'}^n,V^n)\Bigr)_n$ is memoryless,
we also say $(V', V)$ instead of $\Bigl(({V'}^n,V^n)\Bigr)_n$.
Without loss of generality, we assume that $(V', V)$
is binary (since one can easily reduce a non-binary  $(\bar{V'},\bar{V})$ with $I(\bar{V'},\bar{V})>0$ 
to some 
$(V',V)$ with $I(V',V)>0$). The only exception is Section  \ref{lications}, where $(V', V)$
may be not binary.
It has been shown in \cite{Ahl/Cai} that this is a helpful resource for information transmission through an arbitrarily varying classical channel. 
The use of mere correlation  already allows one to transmit messages at any rate that is achievable using the optimal form of shared randomness. 
The capacity of an arbitrarily varying quantum channel assisted by correlated shared randomness as a resource has been discussed in \cite{Bo/No},
 where equivalent results were found.

Our previous work
\cite{Bo/Ca/Ca}  determined the
randomness assisted     capacity with an informed jammer,
where we used randomness as a resource.
However, as \cite{Bo/No} showed,  common randomness  is  a very ``costly'' resource,
we have to require that the sender and the receiver each obtain      pairs of identical signals  that
 a random experiment outputs. 
     Thus in this work we consider the correlation as a resource, which is a much ``cheaper''
resource in the sense that
we can simulate any $(V',V)$
  correlation  by  common randomness asymptotically, however, there exists a class of
sequences of bipartite distributions which cannot model   common randomness (cf. \cite{Bo/No}).
\vspace{0.3cm}

Now we consider the  correlation assisted code.

\begin{definition}\label{lmambfslxyb}  
We assume that the transmitters have access to
an arbitrary correlated source
$(V', V)$ with alphabets $( {\cal V}',  {\cal V})$. 
  A $(V', V)$-correlation assisted
  $(n, J_n)$  code  $\mathcal{C}(V', V)$ for the
arbitrarily varying classical-quantum channel $\{\{\rho(x,s), x \in {\cal X}\}, s \in {\cal S}\}$
consists of a set of  encoders $\left\{u_{{v'}^n}:
\{ 1,\cdots ,J_n\} \rightarrow {{\cal X}}^n:
{v'}^n\in{{\cal V}'}^n\right\}$ and
 a set of collections of positive-semidefinite operators $\Bigl\{\{D_j^{(v^n)}:
  j = 1,\cdots ,J_n\}:v^n\in{\cal V}^n\Bigr\}
$ on ${\cal H}^{\otimes n}$, which fulfills $\sum_{j=1}^{J_n} D_j^{(v^n)} =
\mathrm{id}_{{\cal H}^{\otimes
n}}$ for every ${v}^n\in{\cal V}^n$.\end{definition}
\vspace{0.3cm}

\begin{definition}
\label{lmambfslxyb2}  
Let $(V', V)$ with alphabets $( {\cal V}',  {\cal V})$ be an arbitrary correlated source. 
 A non-negative number $R$ is an achievable  $(V', V)$-correlation assisted rate   
with an informed jammer  
  for the AVCQC $\{\{\rho(x,s), x \in {\cal X}\}, s \in {\cal S}\}$, if for every
$\epsilon>0$, $\delta>0$  and sufficiently large $n$ there
exists a $(V', V)$-correlation assisted   $(n, J_n)$  code
$\mathcal{C}(V', V) =
\biggl\{\Bigl(u_{{v'}^n},\{D_j^{({v}^n)}: j\in\{
1,\cdots ,J_n\}\}\Bigr): {v'}^n\in{{\cal V}'}^n,\text{ }
 v^n\in{\cal V}^n\biggr\}$  such that $\frac{\log J_n}{n}
> R-\delta$ and
\[\max_{ {\bf s}^n (\cdot) }  \sum_{{v'}^n\in{{\cal V}'}^n}\sum_{  v^n\in{\cal V}^n}
p({v'}^n,v^n) P_e(\mathcal{C}({v'}^n,v^n),  {\bf s}^n(\cdot)) < \epsilon\text{ ,}\]
where $P_e(\mathcal{C}({v'}^n,v^n),  {\bf s}^n(\cdot))$  is defined as 
\[ P_e(\mathcal{C}({v'}^n,v^n),{\bf s}^n(\cdot)) := 1- \frac{1}{J_n} \sum_{j=1}^{J_n} 
\mathrm{tr}(\rho(u_{{v'}^n}(j), {\bf s}^n(u_{{v'}^n}(j)))D_j^{(v^n)})\text{ .}\] 

For a given correlated source $(V', V)$,
 the supremum on achievable $(V', V)$-correlation assisted 
   rates  of
 $\{\{\rho(x,s), x \in {\cal X}\}, s \in {\cal S}\}$  with an informed jammer  is called the
 $(V', V)$-correlation assisted    capacity with an informed jammer, denoted by
$C( \{\{\rho(x,s), x \in {\cal X}\}, s \in {\cal S}\};corr(V', V))$.
   Notice that by definition, $C( \{\{\rho(x,s), x \in {\cal X}\}, s \in {\cal S}\};corr(V', V))$
is a function of  $(V', V)$.

 \end{definition}

 \vspace{0.3cm}

In Definition \ref{lmambfslxyb}  and Definition \ref{lmambfslxyb2}  
we assume that  each  channel user obtains one part of a realization
of   $(V', V)$ for every channel use. Now we consider the scenario when
we  constrain the amount of resources we use.

\begin{definition}\label{lmambfslxybwithl}
Let $(V', V)$ with alphabets $( {\cal V}',  {\cal V})$ be an arbitrary correlated source. 
For a sequence of natural numbers $(l_n)_{n\in\mathbb{N}}$,
a $({V'}, V)$-correlation assisted
  $(n, l_n, J_n)$  code  $\mathcal{C}({V'}^{l_n}, V^{l_n})$ for the
arbitrarily varying classical-quantum channel $\{\{\rho(x,s), x \in {\cal X}\}, s \in {\cal S}\}$
consists of a set of  encoders $\left\{u_{{v'}^{l_n}}:
\{ 1,\cdots ,J_n\} \rightarrow {{\cal X}}^n:
{v'}^{l_n}\in{{\cal V}'}^{l_n}\right\}$ and
 a set of collections of positive-semidefinite operators $\Bigl\{\{D_j^{(v^{l_n})}:
  j = 1,\cdots ,J_n\}:v^{l_n}\in{\cal V}^{l_n}\Bigr\}
$ on ${\cal H}^{\otimes n}$, which fulfills $\sum_{j=1}^{J_n} D_j^{(v^{l_n})} =
\mathrm{id}_{{\cal H}^{\otimes
n}}$ for every ${v}^{l_n}\in{\cal V}^{l_n}$.\end{definition}
 \vspace{0.3cm}

\begin{definition}

Let $(V', V)$ with alphabets $( {\cal V}',  {\cal V})$  be an arbitrary correlated source. 
 A non-negative number $R$ is an achievable  $((V', V),(l_n)_{n\in\mathbb{N}})$-correlation assisted rate   
with an informed jammer under the average error criterion 
  for the AVCQC $\{\{\rho(x,s), x \in {\cal X}\}, s \in {\cal S}\}$, if for every
$\epsilon>0$, $\delta>0$ and sufficiently large $n$ there
exists a $(V', V)$-correlation assisted   $(n, l_n,  J_n)$  code
$\mathcal{C}(V', V) =
\biggl\{\Bigl(u_{{v'}^{l_n}},\{D_j^{({v}^{l_n})}: j\in\{
1,\cdots ,J_n\}\}\Bigr): {v'}^{l_n}\in{{\cal V}'}^{l_n},\text{ }
 v^{l_n}\in{\cal V}^{l_n}\biggr\}$  such that $\frac{\log J_n}{n}
> R-\delta$, and
\[\max_{ {\bf s}^n (\cdot) }  \sum_{{v'}^{l_n}\in{{\cal V}'}^{l_n}}\sum_{  v^{l_n}\in{\cal V}^{l_n}}
p({v'}^{l_n},v^{l_n}) P_e(\mathcal{C}({v'}^{l_n},v^{l_n}),  {\bf s}^n(\cdot)) < \epsilon\text{ ,}\]
where $P_e(\mathcal{C}({v'}^{l_n},v^{l_n}),  {\bf s}^n(\cdot))$  is defined as 
\[ P_e(\mathcal{C}({v'}^{l_n},v^{l_n}),{\bf s}^n(\cdot)) := 1- \frac{1}{J_n} \sum_{j=1}^{J_n} 
\mathrm{tr}(\rho(u_{{v'}^{l_n}}(j), {\bf s}^n(u_{{v'}^{l_n}}(j)))D_j^{(v^{l_n})})\text{ .}\]

 The supremum on achievable $((V', V),(l_n)_{n\in\mathbb{N}})$-correlation assisted 
   rates  of
 $\{\{\rho(x,s), x \in {\cal X}\}, s \in {\cal S}\}$  with an informed jammer under the average error criterion is called the
 $((V', V),(l_n)_{n\in\mathbb{N}})$-correlation assisted    capacity with an informed jammer, denoted by
$C( \{\{\rho(x,s), x \in {\cal X}\}, s \in {\cal S}\};corr(V', V),(l_n)_{n\in\mathbb{N}})$.

 \end{definition}
 \vspace{0.3cm}

\section{Main Results and Proofs}\label{mrap}
\subsection{Quantum Version of 
Kiefer and Wolfowitz's  Results for Classical Channels} 

At first, we will      introduce   an important tool
for our main result. That  tool implies the positive 
correlation assisted capacity of an  AVCQC 
when this AVCQC has positive randomness assisted capacity 
and we have access to some correlated source $(V', V)$, such that $I(V', V)>0$.
   
\subsubsection{Proof Concepts} \label{CotP}

Before we go on with our proof for the positivity, we 
would at first like to introduce two
well-known approaches of using
correlation to achieve positive capacity for 
 a classical AVC, and explain 
 why they will not work for
 our model. The first approach
 works       only     for AVC 
(classical or classical-quantum)  when the jammer has no side information,
 but not for AVC when the jammer has  side information.
 The second  approach works       only     for 
classical AVC  when the jammer has  side information,
 but not for AVCQC when the jammer has  side information.

For  classical AVCs,
 \cite{Ahl/Cai} showed 
the equality
of correlation assisted  capacity
and randomness assisted capacity (under the
average error criterion)
for any correlated source $(V', V)$  (with $I(V', V)>0$)
 when the jammer had no side information. 
The idea of the proof 
was at first to show that the correlation assisted  capacity
satisfies the positivity conditions of \cite{Cs/Na}.
Then 
the channel users can create a  sufficient amount of
common randomness  using  codewords of negligible length.
For
this proof it is essential
that the randomness is uniformly distributed.

However, when the jammer has  side information
about the channel input, 
the positivity conditions
of \cite{Cs/Na}
cannot be applied
since there is no Ahlswede Dichotomy  
(cf. \cite{Bo/Ca/Ca}). 
For classical
AVCs with an informed jammer, 
we need to find another
positivity condition.

In \cite{KW62}, Kiefer and Wolfowitz delivered   
      another
condition for the
positivity of a classical AVC (under the
maximal error criterion).
 This is  when we can conclude the classical channel outputs
into two disjunct
convex compact sets in the real vector space.
 When this condition is fulfilled,
then there exists  a hyperplane separating
 the classical channel outputs
 into two parts
in their real vector space.
     The authors of \cite{KW62} 
showed that if there
exists such a hyperplane,
one can construct
a classical binary point to
point channel with positive capacity.


One of the main difficulties is that
we can not apply the classical results of Kiefer and Wolfowitz for 
correlations directly
on the set of quantum states      since they do not form a real vector space.
Thus we have to find a new approach to 
show a quantum version of 
the classical
 positivity condition.

Our idea is concluded in the following two
approaches:
\begin{itemize}
\item At first we
show that if the
the  channel users share
a source $(V', V)$ such that $I(V', V)>0$, then  
there exists  a hyperplane separating
 the quantum outputs
 into two parts.   
\item
 Similar to the classical proposal of 
\cite{KW62}, we will show that when
this condition is satisfied, then the
correlation assisted  capacity of an AVCQC
is 
positive,
when this AVCQC
has 
positive randomness assisted capacity.
\end{itemize}

\subsubsection{A Positivity Condition
for AVCQC
 } 
\label{QVotCPC}

At first we need same notations.

Let $(V', V)$ with alphabets $({\cal V}', {\cal V})$ be an arbitrary correlated source, 
and $\{\rho(x,s):x\in{\cal X}, s\in {\cal S}\}$ be an AVCQC with input alphabet ${\cal X}$ and output space ${\cal H}$. 
For a mapping 
$g: {\cal V}' \rightarrow {\cal X}$ and  a conditional probability distribution $Q \in {\cal P}({\cal S}|{\cal X})$, 
where ${\cal P}({\cal S}|{\cal X})$ is the set of conditional probability distributions from ${\cal X}$ to the state set ${\cal S}$ of ${\cal W}$, we define
\begin{equation}\nu (v|g,Q):= P_{V}(v) P_{V'|V}(g^{-1}(x)|v) \sum_s Q (s|x) \rho\left(x,s\right),\end{equation}
\begin{equation}\vec{\nu} (g,Q):=\left(\nu (v|g,Q)\right)_{v\in V},\end{equation}

and 
\begin{equation} \label{eq_KW02}
{\cal A}(g):= \{\vec{\nu} (g,Q): Q\in{\cal P}({\cal S}|{\cal X})\}.
\end{equation}

For a given AVCQC ${\cal W}=\{\{\rho(x,s), x \in {\cal X}\}, s \in {\cal S}\}$ with set of state ${\cal S}$,  let
\begin{equation} \label{eq_doublebar}
\bar{\bar{{\cal W}}}:=\{\{\bar{\bar{\rho}}_{Q}( x):=\sum_s Q(s|x) \rho(x,s), x \in {\cal X}\}: \mbox{ for all $Q: {\cal X} \rightarrow {\cal S}$}\}.
\end{equation}\vspace{0.5cm}

In the following Lemma \ref{theorema}, we 
show the existence of a hyperplane separating
 the quantum outputs,
 if the
the  channel users share
a source $(V', V)$ such that $I(V', V)>0$.

\begin{lemma} \label{theorema}
    Let $(V', V)$ with alphabets $({\cal V}', {\cal V})$ be an arbitrary correlated source, 
and $\{\rho(x,s):x\in{\cal X}, s\in {\cal S}\}$ 
be an AVCQC with input alphabet ${\cal X}$ and $d$-dimensional output space ${\cal H}$. 
Suppose  $I(V';V)>0$ and $\min_{\rho\in\bar{\bar{{\cal W}}}} C(\rho) >0$ hold.
 Then we can
find $g_0$ and $g_1$ $: {\cal V}' \rightarrow {\cal X}$,  such that
\begin{equation} \label{eq_KW03}
{\cal A}(g_0) \cap {\cal A}(g_1) =\emptyset.
\end{equation}

\end{lemma}\vspace{0.3cm}

{\it Proof:} Let $I(V';V)>0$, $\min_{\rho\in\bar{\bar{{\cal W}}}} C(\rho) >0$, ${\cal V}'={\cal V}=\{0,1\}$ and $|{\cal X}|=\alpha$. We label the input letters in 
${\cal X}$ as $x(0),x(1), \ldots, x(\alpha-1)$. Our proof is based on constructions of two functions 
$g_0: {{\cal V}'}^{\iota} \rightarrow {\cal X}$ and $g_1: {{\cal V}'}^{\iota} \rightarrow {\cal X}$ 
(for a properly defined $\iota$ in the next paragraph), satisfying (\ref{eq_KW03}) and
\begin{equation}\label{eq_a01}
P_{V'}^n(g_0^{-1}(x))=P_{V'}^n(g_1^{-1}(x)),
\end{equation}
for all $x \in {\cal X}$.
Notice that we will need  property (\ref{eq_a01}) for the proof of Theorem \ref{lvvwatcv}.

Let $\iota$ be the smallest integer $\kappa$ such that $\sum_{\tau =0}^{\kappa} \lfloor \frac{1}{2} {\kappa \choose \tau} \rfloor \ge \alpha$.
We shall construct $g_0$ and $g_1$ satisfying (\ref{eq_KW03}). 
To this end we shall group the sequences in ${{\cal V}'}^{\iota}$. But first we need to label them in following way.

$\bullet$ For $h=0,1, \ldots \iota$, divide the sequences in ${{\cal V}'}^{\iota}$ with Hamming weight $h$ into two parts with equal sizes 
$\lfloor \frac{1}{2} {\iota \choose h} \rfloor$ and label them as $u^{\iota}(h,1a), u^{\iota}(h,2a), \ldots, u^{\iota}(h, \lfloor \frac{1}{2} {\iota \choose h} \rfloor_a)$ 
and $u^{\iota}(h,1b), u^{\iota}(h,2b), \ldots, u^{\iota}(h, \lfloor \frac{1}{2} {\iota \choose h} \rfloor_b)$, respectively. 
When ${h \choose \tau}$ is odd, we denote the remaining sequence by $u^{\iota}(h^*)$.

$\bullet$ Order labels $(h,k)$, $h=0,1,2,\ldots, \iota, k=1,2,\ldots, \lfloor \frac{1}{2} {\iota \choose h} \rfloor$
by lexicographic order, as $m=1,2, \ldots, \sum_{\tau =0}^{\iota} \lfloor \frac{1}{2} {\iota \choose \tau} \rfloor$ and rewrite $u^{\iota}(h,k_a)$ 
and $u^{\iota}(h,k_b)$ to $u^{\iota}(m_a)$ and $u^{\iota}(m_b)$ respectively, if $(h,k)$ is the $m$th label in the order.
That is, for $u^{\iota}(m_a)=u^{\iota}(h,k_a)$ and  $u^{\iota}(m'_a)=u^{\iota}(h',k'_a)$,
we have $m_a < m'_a$ if and only if $h <h'$ or $h=h'$ and $k < k'$;
 for $u^{\iota}(m_b)=u^{\iota}(h,k_b)$ and  $u^{\iota}(m'_b)=u^{\iota}(h',k'_b)$, 
we have $m_b < m_b'$ if and only if $h <h'$ or $h=h'$ and $k_a < k'_a$.

Next we assign values of $g_0$ and $g_1$ to the sequences in ${{\cal V}'}^{\iota}$ according to three groups:

{\bf Group 1:} For $m=1,2, \ldots, \alpha-1$, we let $g_0(u^{\iota}(m_a))=g_1(u^{\iota}(m_b))=x(0)$, $g_0(u^{\iota}(m_b))=x(m)$ and $g_1(u^{\iota}(m_a))=x(m)$, 
 respectively. Notice that as $u^{\iota}(m_a)$ and $u^{\iota}(m_b)$ have the same Hamming weight for all $m$,  for every $u^{\iota}\in{{\cal V}'}^{\iota}$, we have
\begin{equation}\label{eq_apr}
Pr\left( g_0( u^{\iota})=  x(m)\right)= Pr\left(u^{\iota}: g_1( u^{\iota})=  x(m)\right)
\end{equation}
 for $m=0,1,\ldots, \alpha-1$ (i. e. for all $x(m) \in  {\cal X}$), 
in the assignment to the members in group 1.

{\bf Group 2:} For all $m=\alpha, \alpha+1, \ldots, \sum_{\tau =0}^{\iota} \lfloor \frac{1}{2} {\iota \choose \tau} \rfloor$, 
we arbitrarily choose $0 \le \zeta_0(m)) < \zeta_1(m)\le \alpha -1$ and let $g_0(u^{\iota}(m_a))= g_1(u^{\iota}(m_b))=x(\zeta_0(m))$ and 
$g_0(u^{\iota}(m_b))= g_1(u^{\iota}(m_a))=x(\zeta_1(m))$. Again, because $u^{\iota}(m_a)$ and $u^{\iota}(m_b)$ have  
the same Hamming weight, (\ref{eq_apr}) also holds for the assignment to the members of group 2.

{\bf Group 3:} Finally for each $u^{\iota}(h^*)$, we arbitrarily choose  a letter in the alphabet ${\cal X}$, say $x(i_h)$, and 
let $g_0(u^{\iota}(h^*))=g_1(u^{\iota}(h^*))=x(i_h)$. Obviously (\ref{eq_apr}) also holds for the assignment of group 3.

Notice that the property (\ref{eq_apr}) of the assignments to the 3 groups yields (\ref{eq_a01}).

Let $\{|v\rangle: v\in{\cal V}\}$ be an orthonormal basis of $\mathbb{C}^{\otimes |{\cal V}|}$. 
Then the output $\vec{v}(g;Q)$ of
the channel can be presented as a classical-quantum state in the $d|{\cal V}|$-dimensional
complex Hilbert space $\mathbb{C}^{\otimes |{\cal V}|}\otimes {\cal H}$:
\begin{equation}\sum_{v} P_V(v)|v\rangle \langle v| 
\otimes P_{V'\mid V} \left(g_i^{-1}(x)\mid v\right)\sum_s Q(s\mid x) \rho(x,s)\text{ ,}\end{equation}
when one applies the function $g_i$, $i = 0, 1$ to the input of the channel according to the
output of the source $V'$. Consequently, in the notation, ${\cal A}(g_i)$ is presented as
$\{\sum_{v} P_V(v)|v\rangle \langle v| 
\otimes P_{V'\mid V} \left(g_i^{-1}(x)\mid v\right)\sum_s Q(s\mid x) \rho(x,s):
 Q \in {\cal P}({\cal S}\mid {\cal X})\}$.

Next we shall show (\ref{eq_KW03}) by assuming a contradiction that (\ref{eq_KW03})
 will not hold. That is, there exist $Q_0, Q_1 \in {\cal P}({\cal S}|{\cal X})$ such that, $\vec{\nu}(g_0,Q_1)=\vec{\nu}(g_1,Q_0)$ or
\[P_V^{\iota}(v^{\iota}) \sum_x P_{V'|V}^{\iota}(g_0^{-1}(x)|v^{\iota}) \sum_s Q_1(s|x) \rho(x,s)=
P_V^{\iota}(v^{\iota}) \sum_{x'} P_{V'|V}^{\iota}(g_1^{-1}(x')|v^{\iota}) \sum_{s'} Q_0(s'|x') \rho(x',s'),\]
for all $v^{\iota}$, which can be rewritten as
\[
P_V^{\iota}(v^{\iota}) \sum_{u^{\iota}} P_{V'|V}^{\iota}(u^{\iota}|v^{\iota})\sum_s Q_1 (s|g_0(u^{\iota})) \rho(g_0(u^{\iota}),s)=
P_V^{\iota}(v^{\iota}) \sum_{u^{\iota}} P_{V'|V}^{\iota}(u^{\iota}|v^{\iota})\sum_{s'} Q_0 (s'|g_1(u^{\iota})) \rho(g_1(u^{\iota}),s'),\]
by re-arranging the terms. That is,
\begin{equation} \label{eq_a02}
\sum_{u^{\iota}} P_{V'|V}^{\iota}(u^{\iota}|v^{\iota})[\sum_s Q_1 (s|g_0(u^{\iota})) \rho(g_0(u^{\iota}),s)-\sum_{s'} Q_0 (s'|g_1(u^{\iota})) \rho(y|g_1(u^{\iota}),s')]=0^{\mathcal{H}},
\end{equation}
for all $v^{\iota}$, where $0^{\mathcal{H}}$ is the zero  ensemble on $\mathcal{H}$.

Denote by $\mathfrak{P}_t$, the $|{\cal V}^t| \times |{{\cal V}'}^t|$ matrix whose $(v,u)$th entry is $P_{V'|V}^t(u^t|v^t)$ for all $t$. 
We write $\mathfrak{P}_1=\mathfrak{P}$. Then we observe that for every positive integer $t$, we have $\mathfrak{P}_t=\mathfrak{P}^{\otimes t}$, i. e. 
$\mathfrak{P}_t$ is $t$th-sensor power of $\mathfrak{P}$. Recall that we have assumed that $I(V';V)>0$, which implies that 
$\frac{P_{V'|V}(0|0)}{P_{V'|V}(0|1)} \not= \frac{P_{V'|V}(1|0)}{P_{V'|V}(1|1)}$, or $\det (\mathfrak{P}) \not=0$.
Therefore $\mathfrak{P}$ is a full rank matrix, and hence so is $\mathfrak{P}_t$ for all $t$. Next let $\vec{w}$ be the $d|{{\cal V}'}^{\iota}|\times d$ matrix,
where $d:=\dim \mathcal{H}$,
 whose components are $\sum_s Q_1 (s|g_0(u^{\iota})) \rho(g_0(u^{\iota}),s)-\sum_{s'} Q_0 (s'|g_1(u^{\iota})) \rho(g_1(u^{\iota}),s'), u^{\iota}\in {{\cal V}'}^{\iota}$, 
in a proper order. Then (\ref{eq_a02}) can be rewritten as
\[
\left(\mathfrak{P}_{\iota}\otimes id^{\mathcal{H}}\right) \vec{w}=0^{\mathcal{H}}.
\]
Because $\mathfrak{P}_{\iota}$ is full rank, the linear function $\mathfrak{P}_{\iota} z^{|{{\cal V}'}|^{\iota}}=0$ with respect 
to $z^{|{{\cal V}'}|^{\iota}}$ has no non-zero solution. This implies that $\vec{w}={0^{\mathcal{H}}}^{\otimes|{{\cal V}'}|^{\iota}} $, or
\begin{equation} \label{eq_a03}
\sum_s Q_1 (s|g_0(u^{\iota})) \rho(g_0(u^{\iota}),s)=\sum_{s'} Q_0 (s'|g_1(u^{\iota})) \rho(g_1(u^{\iota}),s'),
\end{equation}
for all $u^{\iota}$. Now  we substitute $u^{\iota}(m_a)$ for $m=1,2 \ldots, \alpha-1$ in the group 1 to (\ref{eq_a03}) and then have that
\[\sum_s Q_1 (s|x(0)) \rho(x(0),s)=\sum_{s'} Q_0 (s'|x(m)) \rho(x(m),s')\]
for $m=1,2, \ldots \alpha-1$. By choosing $(\sum_s Q_1 (s|x(0)) \rho(x(0),s))$ and
 $(\sum_{s'} Q_0 (s'|x(m)) \rho(x(m),s'))$ for $m=1,2, \ldots, \alpha-1$, 
we have a  channel in $\bar{\bar{{\cal W}}}$ with an identity row, 
which contradicts the assumption $\min_{\rho\in\bar{\bar{{\cal W}}}} C(\rho) >0$, and 
therefore (\ref{eq_KW03}) is proven. 

\qedwhite\vspace{0.5cm}

In the following Lemma, we 
show that this condition is
indeed a ``positivity condition'',
i.e. this implies a positive
correlation assisted  capacity.

\begin{lemma} \label{lemmaKW}
There is a positive $r$ such that for all $\epsilon, \lambda >0$, there is an $(n, 2^{nr})$ correlation assisted code for a
sufficiently large $n$, with maximum probability of error 
smaller than $\lambda$ and rate $r'> r-\epsilon$, if there exist two mappings $g_k, k=0,1$ from $V'$ to ${\cal X}$ with
\begin{equation} \label{eq_KW03a}
{\cal A}(g_0) \cap {\cal A}(g_1) =\emptyset.
\end{equation}
\end{lemma}\vspace{0.3cm}

{\it Proof:} We show the lemma in the same way as in \cite{KW62}.

The set $P(\mathcal{S}|\mathcal{X})$ 
is bounded and equal to $\overline{P(\mathcal{S}|\mathcal{X})}$ 
and thus convex and compact.
For a fixed $g$, $\vec{v}(g,\cdot)$ $:P(\mathcal{S}|\mathcal{X})
\rightarrow \mathcal{H}^{|\tilde{\mathcal{V}}|}$ is linear, thus
for every given $g$, $\mathcal{A}(g)$ is also a convex compact set.

 ${\cal A} (g_k), k=0,1$ are compact convex sets in the $|{\cal V}||\mathcal{X}|$-dimensional real space.
When  $\mathcal{A}(g_0) \cap \mathcal{A}(g_1) = \emptyset$ holds,
then there is a hyperplane $\grave{H}$ which separates 
$\mathcal{A}(g_0)$ and $\mathcal{A}(g_1)$. Let $b$ be a
real number such that $\grave{H}-b$ is a subspace.
By Riesz's representation theorem  we can find a self-adjoint operator $A'$ 
$=\Bigl(a(v)\Bigr)_{v \in {\cal V}}$ on ${\cal H}^{d|\tilde{\mathcal{V}}| }$ such
that \[\mathrm{tr}\left((A'-b\cdot\mathrm{id})(\vec{v}(g_0,s))\right)>0\] for all $\vec{v}(g_0,s) \in \mathcal{A}(g_0)$
and \[\mathrm{tr}\left((A'-b\cdot\mathrm{id})(\vec{v}(g_1,s))\right)<0\] for all $\vec{v}(g_1,s) \in \mathcal{A}(g_1)$.

We define the
the self-adjoint operator $A$ on ${\cal H}^{d|\tilde{\mathcal{V}}|}$ 
by $A:= A'-b\cdot\mathrm{id}$.
Suppose $A$ has a  spectral decomposition $A=\sum_{l=1}^{d|\mathcal{V}|}a_l A_l$.
Similar to \cite{Ahl/Bj/Bo/No},
we define for every $m\in\mathbb{N}$ and $l^m=(l_1,\cdots,l_m)$
\[\mathsf{P}_0^m:= \sum_{l^m: \sum_{i=1}^m a_{l_i}<0}  A_{l^m}\text{;}~~
\mathsf{P}_1^m:= \sum_{l^m: \sum_{i=1}^m a_{l_i}>0} A_{l^m}\text{ ,}\]
where $A_{l^m} :=\bigotimes_{i=1}^m A_{l_i}$. We have  $\mathsf{P}_0 + \mathsf{P}_1= \mathrm{id}^m$.

We define
\[C:=\max_{\vec{v}(j,Q) \in \mathcal{A}(g_j)}
\frac{1}{4}\mathrm{tr}(A \vec{v}(g_j,Q))^{-2}
\left(\mathrm{tr}(A^2 \vec{v}(g_j,Q))
-\mathrm{tr}(A \vec{v}(g_j,Q))^2
\right)
.  \]



Now we consider every term in  $\sum_{v^m} P_{V}^m(v^m) \sum_{x^m} P_{V'|V}^m((g_0^{-1})^{m}(x^m)|v^m)\sum_{s^m} Q^m(s^m|x^m)
\rho^{\otimes m}(x^m, {s}^m)$
$=\sum_{v^m} P_{V}^m(v^m)$ $\sum_{u^m} P_{V'|V}^m(u^m|v^m)$ $\sum_{s^m} Q^m(s^m|g_0^{m}(u^m))$
$\rho^{\otimes m}(g_0^{m}(u^m), {s}^m)$. 
We define 
\begin{align*}&L^m:= \biggl\{l^m: \left|\sum_{i=1}^m a_{l_i}
- \min_{Q \in {\cal P}({\cal S}|{\cal X})}\mathrm{tr}\left( A\left[
P_{V}(v_i) P_{V'|V}(g_0^{-1}(x)|v_i) \sum_s Q (s|x) \rho\left(x,s\right)\right]
\right)\right|\\
&\leq\frac{1}{2} \min_{Q \in {\cal P}({\cal S}|{\cal X})}\mathrm{tr}\left( A\left[
P_{V}(v_i) P_{V'|V}(g_0^{-1}(x)|v_i) \sum_s Q (s|x) \rho\left(x,s\right)\right]
\right)\biggr\}.\end{align*}
For 
every fixed $u\in \mathcal{U}$,
we have $\{\mathbf{s}(g_0(u)) , \mathbf{s}(\cdot) \in \mathbf{S} \}$ $=\mathcal{S}$,
thus similar to the proof in  \cite{Ahl/Bj/Bo/No} we may
apply Chebyshev's inequality for every $s^m\in \mathcal{S}$ to show

\begin{align}&\min_{\mathbf{s}^m(\cdot)\in  \mathbf{S}}\mathrm{tr}\left(\mathsf{P}_0^m\left(P(v^m)\rho (g_0^m(v^m),\mathbf{s}^m(g_0^m(v^m)))\right)_{v^m\in{\cal V}^m}\right)\allowdisplaybreaks\notag\\
&=\min_{\mathbf{s}^m(\cdot) \in  \mathbf{S}}
\mathrm{tr}\left(\left(\sum_{l^m: \sum_{i=1}^m a_{l_i}<0}  A_{l^m}\right)\left(P_{V}(v^m)\rho (g_0^m(v^m),\mathbf{s}^m(g_0^m(v^m)))\right)_{v^m\in{\cal V}^m}\right)\allowdisplaybreaks\notag\\
&=\min_{\mathbf{s}^m(\cdot) \in  \mathbf{S}}\sum_{l^m: \sum_{i=1}^m a_{l_i}<0} \prod_{i=1}^{m} \mathrm{tr}\left( A_{l_i}
\left(P_{V}(v_i)\rho (g_0(v_i),\mathbf{s}_i(g_0(v_i)))\right)\right)\allowdisplaybreaks\notag\\
&\geq\min_{\mathbf{s}^m(\cdot) \in  \mathbf{S}}\sum_{l^m\in L^m} \prod_{i=1}^{m} \mathrm{tr}\left( A_{l_i}
\left(P(v_i)\rho (g_0(v_i),\mathbf{s}_i(g_0(v_i)))\right)\right)\allowdisplaybreaks\notag\\
&\geq\sum_{l^m\in L^m} \prod_{i=1}^{m} \min_{Q \in {\cal P}({\cal S}|{\cal X})}\mathrm{tr}\left( A_{l_i}\left[
P_{V}(v_i) P_{V'|V}(g_0^{-1}(x)|v_i) \sum_s Q (s|x) \rho\left(x,s\right)\right]
\right)\allowdisplaybreaks\notag\\
&\geq 1-\frac{1}{4m}\max_{Q \in {\cal P}({\cal S}|{\cal X})}\biggl[
(\mathrm{tr}(A P_{V}(v_i) P_{V'|V}(g_0^{-1}(x)|v_i) \sum_s Q (s|x) \rho\left(x,s\right)))^{-2}\allowdisplaybreaks\notag\\
&\left(\mathrm{tr}(A^2 P_{V}(v_i) P_{V'|V}(g_0^{-1}(x)|v_i) \sum_s Q (s|x) \rho\left(x,s\right))
-\mathrm{tr}(A P_{V}(v_i) P_{V'|V}(g_0^{-1}(x)|v_i) \sum_s Q (s|x) \rho\left(x,s\right))^2\right)\biggr]\allowdisplaybreaks\notag\\
&= 1-\frac{1}{m}\max_{\vec{v}(g_0,Q) \in \mathcal{A}(g_0)}
\frac{1}{4}\mathrm{tr}(A \vec{v}(g_0,Q))^{-2}
\left(\mathrm{tr}(A^2 \vec{v}(g_0,Q))
-\mathrm{tr}(A \vec{v}(g_0,Q))^2
\right)\allowdisplaybreaks\notag\\
&\geq 1-\frac{C}{m},\end{align}
where $N(s|s^m)$ the number of occurrences of the symbol $s$ in $s^m$.\vspace{0.2cm}

Similarly,
 \begin{equation}\min_{\mathbf{s}^m(\cdot)\in  \mathbf{S}}\mathrm{tr}\left(\mathsf{P}_1^m\left(P(v^m)\rho (g_0^m(v^m),\mathbf{s}^m(g_0^m(v^m)))\right)_{v^m\in{\cal V}^m}\right)
\geq 1-\frac{C}{m}.\end{equation} \vspace{0.4cm}

Using the idea of  \cite{Ahl/Bj/Bo/No} we can now define a classical binary
AVC by 
\[\hat{W}(0|g_i,s^m):=\sum_{(v^m,y^m) \in {\cal B}} \sum_{x^m} P_{V}^m(v^m) P_{V'|V}^m((g_i^{m})^{-1}(x^m)|v^m)\rho^m(x^m,\mathbf{s}^m(x^m))\]
and
\[\hat{W}(1|g_i,s^m):=\sum_{(v^m,y^m) \in {\cal B}^c} \sum_{x^m} P_{V}^m(v^m) P_{V'|V}^m((g_i^{m})^{-1}(x^m)|v^m)\rho^m(x^m,\mathbf{s}^m(x^m))\] for $i=0,1$,
\[\hat{\cal W}:=\{\hat{W}(\cdot|\cdot, \hat{s}), \hat{s} \in \hat{\cal S}\} \mbox{ for $\hat{\cal S}=\mathbf{S}^m$}.\]
$\hat{\cal W}$ (with input alphabet $\{g_0^m,g_1^m\}$) is a binary AVC such that $\hat{W}(0|g_0,\hat{s})>1-\eta$ and 
$\hat{W}(0|g_1,\hat{s'})<\eta$ for all $\hat{s}, \hat{s'} \in \hat{\cal S}$.
 Every deterministic code for $\hat{W}$ also defines a correlation assisted code for $\rho$.

Now we apply a lemma from Ahlswede and Wolfowitz \cite{AW70}.

\begin{lemma}   Let $W^{bc} := \left\{\{W(j\mid i, s), i, j \in \{0,1\}\}, s\in{\cal S}\right\}$ be a binary
classical AVC. Then its deterministic code capacity is equal to 
\[\max_P \min_{\bar{\bar{W}}\in \bar{\bar{W^{bc}}}} I(P, \bar{\bar{W}})>0, \]
 if for all $s,s'$, we have $W(0\mid 0,s)  + W(1\mid 1,s') > 1$.\label{mpmbbwibbwbc}
\end{lemma}

Applying Lemma \ref{mpmbbwibbwbc}
we can construct a  binary point to
point channel with positive capacity.
This shows 
Lemma \ref{lemmaKW}.
\qedwhite\vspace{0.5cm}

\subsubsection{An Alternative Proof} 
\label{APaaPftAAVClC} 

In Section \ref{CotP}, we point out that the concept of
\cite{KW62} would work for classical AVC with an informed  jammer, but not for
AVCQC with an informed  jammer. In
this section we would like to introduce
a ``trick''  to apply
the concept of
\cite{KW62} on AVCQC. This delivers an
alternative proof for Lemma \ref{lemmaKW}.

At first, observe that  as mentioned in Section \ref{CotP}, the reason why
we can not apply the classical
results  of 
\cite{KW62} on the set of quantum states
$\{\nu (v|g,Q)\}$ directly is that the set of the probability
matrices do not form a real vector space over $\mathbb{C}$.
However, there is an isomorphism  which maps every set of
Hermitian complex $m\times m$ matrices to a 
$m^2$-dimensional 
subspace of the $2m^2$-dimensional vector space of complex
$m\times m$  matrices over  $\mathbb{R}$. Thus
alternatively, we can  prove 
Lemma \ref{lemmaKW} as follows.

Let $\{|v\rangle: v\in{\cal V}\}$ be an orthonormal basis of ${\cal H}^{\otimes |{\cal V}|}$. 
As mentioned in the proof of Lemma \ref{theorema} above,
the channel can be presented as a classical-quantum state in the $d|{\cal V}|$-dimensional
complex Hilbert space:
\begin{equation}\sigma_{Q,g_i}:= \sum_{v} P_V(v)|v\rangle \langle v| 
\otimes P_{V'\mid V} \left(g_i^{-1}(x)\mid v\right)\sum_s Q(s\mid x) \rho(x,s)\text{ ,}\end{equation}
when one applies the function $g_i$, $i = 0, 1$ to the input of the channel according to the
output of the source $V'$.  ${\cal A}(g_i)$, presented as
$\{\sigma_{Q,g_i}: Q \in {\cal P}({\cal S}\mid {\cal X})\}$, is a compact convex subset in the 
real vector space $\mathfrak{S}$ formed
by $d|{\cal V}| \times d|{\cal V}|$ Hermitian matrices, a $(d|{\cal V}|)^2$-dimensional real Hilbert subspace of
$2(d|{\cal V}|)^2$-dimensional space of complex $d|{\cal V}| \times d|{\cal V}|$ 
 matrices (with the inner product of
$A$ and $B$, $\langle A | B \rangle = tr(A*B)$).

Let $\phi$ be a linear isomorphic mapping from $\mathfrak{S}$ to
the $(d|{\cal V}|)^2$-dimensional real linear vector space $\mathfrak{V}$, (keeping the inner product unchanged
 $\langle A | B \rangle = tr(A*B)$). 
     Then $\phi(A(g_i)) := \{\phi(\sigma_{Q,g_i}) : Q \in {\cal P}({\cal S}\mid {\cal X})\}$
for $i = 0, 1$ are
compact convex subsets in $\mathfrak{V}$. 
By Lemma \ref{theorema} we have
\[{\cal A}(g_0) \cap {\cal A}(g_1) =\emptyset.\]
Then we have
\[\phi({\cal A}(g_0)) \cap \phi({\cal A}(g_1)) =\emptyset\]
as well, by the isomorphism. Thus, $\phi({\cal A}(g_0))$ and $\phi({\cal A}(g_0))$ can be separated by a
hyperplane. Namely, there is a $(d|{\cal V}|)^2$-dimensional real vector $\vec{\mathbf{a}}$
 and a real number
$b$ such that
\[\langle \phi(\sigma_{Q_0,g_0}) | \vec{\mathbf{a}} \rangle   < b <\langle \phi(\sigma_{Q_1,g_1}) | \vec{\mathbf{a}} \rangle \]
 for all  $\phi(\sigma_{Q_i,g_i}) \in \phi(A(g_i))$, $i = 0, 1$.
Let $A$ be the inverse image of  $\vec{\mathbf{a}}$ under the mapping $\phi$. Then by the isomorphism
again, we have that
\begin{equation} tr(\sigma_{Q_0,g_0}A) < b < tr(\sigma_{Q_1,g_1}A)\label{tsq0g0a}\end{equation}
 for all  $\sigma_{Q_i,g_i} \in A(g_i)$, $i = 0, 1$.
(Notice, we have that  $\sigma_{Q_i,g_i}* =  \sigma_{Q_i,g_i}$
and $A* = A$ here).
We can now 
apply the classical
results of
Kiefer and Wolfowitz
to construct a  binary point to
point classical channel with positive capacity.
 This delivers an
alternative  proof for
Lemma \ref{lemmaKW}.
\vspace{0.5cm}

\subsection{Correlation Assisted  Capacity Formula}
\label{CACF}

 Our next step is creating a  sufficient amount of
common randomness,
     similar to  the technique in \cite{Bj/Bo/Ja/No} and
\cite{Bo/Ca/De2}.
 In our previous work \cite{Bo/Ca/Ca}, we delivered 
the randomness assisted capacity 
 when the jammer has side information about
the channel input. 
For this proof, only a negligible amount
of randomness was needed.
This, together with our last step, demonstrates
the equality
of correlation assisted  capacity
and randomness assisted capacity
for  AVCQCs. For
our proof it is essential
that the  randomness we create in the last step must be uniformly distributed.

\begin{theorem}
Let $(V', V)$ with alphabets $({\cal V}', {\cal V})$ be an arbitrary correlated source 
and ${\cal W}=\{\rho(x,s):x\in{\cal X}, s\in {\cal S}\}$ be an AVCQC. 
When $I(V', V)>0$ holds, then
 \begin{equation}C( {\cal W};corr(V', V))=\max_P  \min_{\bar{\bar{\rho}}(\cdot) \in \bar{\bar{W}}} 
\chi(P, \bar{\bar{\rho}}(\cdot)). \end{equation} Here
$\bar{\bar{{\cal W}}}$ is defined as in 
(\ref{eq_doublebar}), i.e.
$\{\{\bar{\bar{\rho}}_{Q}( x):=\sum_s Q(s|x) \rho(x,s), x \in {\cal X}\}: \mbox{ for all $Q: {\cal X} \rightarrow {\cal S}$}\}$.
\label{lvvwatcv}
\end{theorem}

 Our model is a generalization of the
standard AVC model, since  the standard ACV model
is limited to the case when the jammer has no side information about the codeword.
Furthermore, our model is more complicated 
than the standard AVC model.
Since the jammer can choose his
jamming strategy according to a block of the channel input, the size of the eavesdropper's possible
output will be a double-exponential of the code length.
    Notice that when we have no access to any resource, then
the (deterministic) capacity formula
for arbitrarily varying channels with
an informed jammer is still an open problem,
even for classical arbitrarily varying channels.
In classical information theory it is known
that zero error capacity can be reduced to a special case (cf. \cite{Bo/Ca/Ca} and \cite{Ahl0}). 
However, to determine zero error capacity is one of
the hardest well-known open problems of more than six
decades, since 1956 (cf. \cite{Sha}).
Theorem \ref{lvvwatcv} delivers a single letter characterization
of the correlation assisted capacity with an informed jammer.
This
is particularity interesting and
promising,  since it shows that correlation, the
 weakest  form of resource,  
is already powerful enough to protect
against such a mighty jamming strategy, and
is as helpful
as common randomness, the ``costly'' resource.

{\it Proof:}
In \cite{Bo/Ca/Ca} it has been shown that the randomness
assisted  capacity of
 $ {\cal W}$
under the average error criterion with an informed jammer
is equal to $\min_{\bar{\bar{\rho}}(\cdot) \in \bar{\bar{W}}} 
\chi(P_X, \bar{\bar{\rho}}(\cdot))$.
In this proof a we use a random variable
uniformly distributed on a  finite set $\mathcal{K}$ 
 such that 
\[\left|\mathcal{K}\right|= c_kn^2,\]
where $c_k$ is a  positive constant depending on $|{\cal X}|$ and $|{\cal S}|$.
\vspace{0.2cm}

Since the  $(V', V)$-correlation assisted    capacity cannot
exceed the randomness assisted    capacity, the converse is trivial.

When $\min_{\bar{\bar{\rho}}(\cdot) \in \bar{\bar{W}}} 
\chi(P_X, \bar{\bar{\rho}}(\cdot))$ $=0$ holds, then the  randomness assisted    capacity
of  ${\cal W}$ is equal to zero and thus  the  $(V', V)$-correlation assisted    capacity
of  ${\cal W}$ is also equal to zero. This case is  trivial. Now
we assume that both $I(V', V)>0$ and $\min_{\bar{\bar{\rho}}(\cdot) \in \bar{\bar{W}}} 
\chi(P_X, \bar{\bar{\rho}}(\cdot))$ $>0$ hold.
Our idea  is to build   a two-part code word, the first part is used to
create the common randomness for the sender and the  receiver, the second is used to transmit the message to the 
receiver.
\vspace{0.2cm}

Our idea is at first to build a pre-code, which is a $(V', V)$-correlation assisted code,  generating
the random variable
uniformly distributed on a finite set $\mathcal{K}$ we used for the randomness assisted code in
\cite{Bo/Ca/Ca}.
 The next step is to
apply the  result of \cite{Bo/Ca/Ca}, i.e. a
second code for message transmission, which is 
a randomness
 assisted code using the random variable we generated.
Thus the codeword  we use at the end is a
two-part codeword.  \vspace{0.2cm}

\bf Definition of pre-code \rm\vspace{0.2cm}

Let $\mathcal{K}$ be a finite set such that $\left|\mathcal{K}\right|$ is of polynomially size of $n$.
We denote $\nu(n):= \frac{3}{r}\log n$, where $r$ is defined as in
Lemma \ref{lemmaKW}. Recall that $\left|\mathcal{K}\right|= c_kn^2<2^{\nu(n)r}$.

 By  Lemma \ref{lemmaKW} we can apply the coding theorem of AVCs with a binary output.
Every deterministic code for $\hat{W}$ also defines a correlation assisted code for $\rho$. Thus by 
Lemma \ref{lemmaKW} there exists a
 $(V', V)$-correlation assisted   code
$\Bigl(\left(u_{{v'}^{\nu(n)}}(k)\right)_{k=1,\cdots,\left|\mathcal{K}\right|},   \{D_{k}^{\nu(n)}:
k=1,\cdots,\left|\mathcal{K}\right|\}\Bigr)$ with deterministic encoder $u_{{v'}^{\nu(n)}}(k) \in\{g_0^m,g_1^m\}^{\nu(n)}$
for ${\cal W}$ of
length $\nu(n)$ (cf. \cite{KW62}),
such that
\begin{align} &
\min_{s^{\nu(n)}(\cdot)} \frac{1}{\left|\mathcal{K}\right|}
 \sum_{k=1}^{\left|\mathcal{K}\right|}
\mathrm{tr}\biggl(\rho(u_{{v'}^{\nu(n)}}(k), s^{\nu(n)}(u_{{v'}^{\nu(n)}}(k)))
D_{k}^{\nu(n)}\biggr)\geq 1-\vartheta \text{ .} \label{ineqfornocausal}\end{align} 

Here we use the property (\ref{eq_a01}) in the proof of Lemma
\ref{theorema}, that $P_{V'}^n(g_0^{-1}(x))=P_{V'}^n(g_1^{-1}(x))$
for all $x \in {\cal X}$,  to show that the randomness we built 
is a uniformly distributed random variable.\vspace{0.2cm}

When the 
 jammer  knew the shared randomness, he could
render the shared randomness completely useless (cf. \cite{Bo/Ca/De3}).
Thus we have to guarantee that 
 the jammer, knowing the code word, has no access to  the randomness. 
Notice that the sender sends the input $(g_{{v'}_1},g_{{v'}_2}, \ldots, g_{{v'}_{\nu(n)}})$ of the AVCQC ${\cal W}$, 
if he wants to send a binary input $({v'}_1{v'}_2 \ldots {v'}_{\nu(n)})$ to the binary classical AVC $\hat{W}$.
Since $P_{V'}^n(g_0^{-1}(x))=P_{V'}^n(g_1^{-1}(x))$
holds, for any  $x^{\nu(n)}\in\mathcal{X}^{\nu(n)}$ we have
\[Pr\left(g_{{{v'}_1}^{\nu(n)}}^{-1}(x^{\nu(n)})\right)=Pr\left(g_{{v_2'}^{\nu(n)}}^{-1}(x^{\nu(n)})\right), \]
for every ${v_1'}^{\nu(n)}$, ${v_2'}^{\nu(n)}$ $\in {\tilde{\cal V'}}^{\nu(n)}$.
This means that
the jammer, knowing the code word, has no access to the randomness.

\vspace{0.4cm}

\bf Definition of the two-part code \rm\vspace{0.2cm}

By \cite{Bo/Ca/Ca} there is an $(n , J_n)$ random code
$\Bigl\{u(j,k ), D_{k,j}^{n}\Bigr\}$, such
that
\[\max_{s(\cdot)}\frac{1}{|{\cal J}|} \frac{1}{\left|\mathcal{K}\right|}
\sum_{j \in {\cal J}} \sum_{k \in {\cal K}} 
tr[\rho^{\otimes n}(u(j,k), s (u(j,k)))(\mathbb{I}_{\cal H}-{\cal D}(j,k))]<\lambda .\]\vspace{0.2cm}

Now we can construct a $(V',V)$-correlation assisted $(\nu(n) +n,
J_n)$ random code $\mathcal{C}(V',V) =
\biggl\{\Bigl(u_{{v'}^{\nu(n)+n}},\{D_j^{v^{\nu(n)+n}}:
j\in \{ 1,\cdots
,J_n\}\}\Bigr) :{v'}^{\nu(n)+n}\in{\mathcal{V}'}^{\nu(n)+n},v^{\nu(n)+n}\in\mathcal{V}^{\nu(n)+n}\biggr\}$,
where we set
\[u_{{v'}^{\nu(n)+n}}(j,k)=
(u_{{v'}^{\nu(n)}}(k) ,u(j,k))
\text{ ,}
\] and \[D_j^{v^{\nu(n)+n}} := \sum_{k=1}^{\left|\mathcal{K}\right|}
D_{(v^{\nu(n)}),k}^{\nu(n)}\otimes D_{k,j}^{n} \text{ .}
\]

\begin{remark}
Notice that the first part of this two-part codeword does not
depend on the message, while the second part does not depend
on the correlation. Thus our $(n,J_n)$ code $\mathcal{C}(V',V)$
is actually a $(\nu(n)+n,J_n, \nu(n))$ code
of $\nu(n)+n$ length.
\label{nttfpot}
\end{remark}\vspace{0.2cm}

By (\ref{ineqfornocausal}), for any $s^{\nu(n) +n}(\cdot)$ we have
\begin{align}&1-\sum_{{v'}^{\nu(n) +n}} \sum_{v^{\nu(n) +n}}
p({v'}^{\nu(n) +n},v^{\nu(n)
+n})\frac{1}{J_n}\sum_{j=1}^{J_n}\mathrm{tr}\biggl(\biggl[\frac{1}{\left|\mathcal{K}\right|}
 \sum_{k=1}^{\left|\mathcal{K}\right|}\allowdisplaybreaks\notag\\
&\rho(u_{{v'}^{\nu(n)}}(k), s^{\nu(n)}(u^{\nu(n)}))\otimes \rho(u(j,k),s^{n}(u(j,k))))\biggr]\cdot\left[\sum_{k=1}^{\left|\mathcal{K}\right|}
D_{(v^{\nu(n)}),k}^{\nu(n)}\otimes D_{k,j}^{n}\right]
\biggr)\allowdisplaybreaks\notag\\
&\leq 1-\sum_{{v'}^{\nu(n) +n}} \sum_{v^{\nu(n) +n}}
p({v'}^{\nu(n) +n},v^{\nu(n)
+n})
\frac{1}{J_n}\sum_{j=1}^{J_n}\mathrm{tr}\biggl(\frac{1}{\left|\mathcal{K}\right|}
 \sum_{k=1}^{\left|\mathcal{K}\right|}\allowdisplaybreaks\notag\\
&\left[\rho(u_{{v'}^{\nu(n)}}(k), s^{\nu(n)}(x^{\nu(n)}))\otimes
  \rho(u(j,k),s^{n}(u(j,k)))\right]\cdot\left[
D_{(v^{\nu(n)}),k}^{\nu(n)}\otimes D_{k,j}^{n}\right]
\biggr)\allowdisplaybreaks\notag\\
&=1-\sum_{{v'}^{\nu(n) }} \sum_{v^{\nu(n) }}
p({v'}^{\nu(n) },v^{\nu(n)})
\frac{1}{J_n}\sum_{j=1}^{J_n}\mathrm{tr}\biggl(\frac{1}{\left|\mathcal{K}\right|}
 \sum_{k=1}^{\left|\mathcal{K}\right|}\allowdisplaybreaks\notag\\
&\left[\rho(u_{{v'}^{\nu(n)}}(k), s^{\nu(n)}(u_{{v'}^{\nu(n)}}(k)))
 D_{(v^{\nu(n)}),k}^{\nu(n)}\right]\otimes\left[\sum_x
\rho(u(j,k),s^{n}(u(j,k))) D_{k,j}^{n}\right]
\biggr)\allowdisplaybreaks\notag\\
&=1-\sum_{{v'}^{\nu(n) }} \sum_{v^{\nu(n) }}
p({v'}^{\nu(n) },v^{\nu(n)})\frac{1}{\left|\mathcal{K}\right|}
 \sum_{k=1}^{\left|\mathcal{K}\right|}\mathrm{tr}\left(
\rho(u_{{v'}^{\nu(n)}}(k),s^{\nu(n)}(u_{{v'}^{\nu(n)}}(k)))
 D_{(v^{\nu(n)}),k}^{\nu(n)}\right) \allowdisplaybreaks\notag\\
&\cdot\left(\frac{1}{J_n}\sum_{j=1}^{J_n} \mathrm{tr}(
\rho(u(j,k),s^{n}(u(j,k)^{n})) D_{k,j}^{n})\right)\allowdisplaybreaks\notag\\
&\leq \lambda+\vartheta\text{ .}\label{errordetb}\end{align}\qedwhite
\vspace{0.5cm}

In Theorem \ref{lvvwatcv}
we consider that for every channel use,
the  channel users receive one signal per channel
use from the correlation.
Since the correlation, which can be regard as some
 ``public communication'' may have a cost, 
we want to analyze now the scenario when,
instead of one signal per channel use,
only  a small number of signals from the correlation
for mutable channel uses are available.

\begin{corollary}
Let $(V', V)$ with alphabets $({\cal V}', {\cal V})$ be an arbitrary correlated source 
and ${\cal W}=\{\rho(x,s):x\in{\cal X}, s\in {\cal S}\}$ be an AVCQC. 
There is a positive $r''$ such that
    for any  sequence of natural numbers $(l_n)_{n\in\mathbb{N}}$ such that $r''<$ $\liminf_{n\rightarrow\infty} \frac{l_n}{\log n}$
$\leq\limsup_{n\rightarrow\infty} \frac{l_n}{\log n}$ $< \infty$,
   when  $I(V', V)>0$ holds, we have
 \begin{equation}C( {\cal W};corr(V', V),(l_n)_{n\in\mathbb{N}})= \min_{\bar{\bar{\rho}}(\cdot) \in \bar{\bar{W}}} 
\chi(P_X, \bar{\bar{\rho}}(\cdot)).\end{equation} 

\label{awcvvlmrc}
\end{corollary}
{\it Proof:}      We define $r'':= \frac{3}{r}$,
where $r$ is defined as in Lemma \ref{lemmaKW}.
Let $(l_n)_{n\in\mathbb{N}}$ be a sequence such that
$\frac{\nu(n)}{\log n}$ $=r''$ $<\liminf_{n\rightarrow\infty} \frac{l_n}{\log n}$
$\leq\limsup_{n\rightarrow\infty} \frac{l_n}{\log n}$ $< \infty$.
We have: $l_n \geq \nu(n)$ for all $n$.
     By Remark \ref{nttfpot}, for any positive $\delta$, $\epsilon$ and sufficiently large $n$ there 
is  a $(l_n+n,J_n, l_n)$ code,
where $J_n = 2^{n(\min_{\bar{\bar{\rho}}(\cdot) \in \bar{\bar{W}}}\chi(P_X, \bar{\bar{\rho}}(\cdot)) - \delta)}$,
such that 
\[\max_{ {\bf s}^{l_n+n} \in \mathbf{S}^{l_n+n}}  \sum_{{v'}^{l_n}\in{{\cal V}'}^{l_n}}\sum_{  v^{l_n}\in{\cal V}^{l_n}}
p({v'}^{l_n},v^{l_n}) P_e(\mathcal{C}({v'}^{l_n},v^{l_n}),  {\bf s}^{l_n+n}) < \epsilon\text{ .}\]
 Since 
$2^{l_n}$ is in polynomial order of $n$,
for any positive $\varepsilon$, if $n$ is large enough, we have $\frac{1}{n}\log J_n -\frac{1}{l_n +n}\log J_n
\leq \varepsilon$.
Thus when $I(V', V)>0$,
\[C( {\cal W};corr(V', V),(l_n)_{n\in\mathbb{N}})= \min_{\bar{\bar{\rho}}(\cdot) \in \bar{\bar{W}}} 
\chi(P_X, \bar{\bar{\rho}}(\cdot)).\] \qedwhite\vspace{0.5cm}

In Theorem \ref{lvvwatcv}
we consider the scenario when
the positivity condition of Lemma \ref{theorema}
is fulfilled.
What remains is to analyze the scenario when
this positivity condition does not hold.

\begin{lemma} Let
 ${\cal W}=\{\rho(x,s):x\in{\cal X}, s\in {\cal S}\}$ be an AVCQC. 
When for every $n\in \mathbb{N}$ and every $x_1$, $x_2\in {\cal X}^n$,
\[conv\left(\{\rho^{\otimes n} (x_1,s^n): s^n\in {\cal S}^n\} \right) \cap conv\left(\{\rho^{\otimes n} (x_1,s^n): s^n\in {\cal S}^n\} \right) \not= \emptyset \]
holds, then the deterministic capacity of ${\cal W}$ with an informed jammer
is zero.
\end{lemma} 
{\it Proof:} 
When for every $n\in \mathbb{N}$ and every $x_1$, $x_2\in {\cal X}^n$,
\[conv\left(\{\rho^{\otimes n} (x_1,s^n): s^n\in {\cal S}^n\} \right) \cap conv\left(\{\rho^{\otimes n} (x_1,s^n): s^n\in {\cal S}^n\} \right) \not= \emptyset \]
holds, by \cite{Ahl/Bj/Bo/No} the deterministic capacity of ${\cal W}$ with an uninformed jammer
is zero.
Since the deterministic capacity of ${\cal W}$ with an informed jammer
cannot exceed the deterministic capacity of ${\cal W}$ with an uninformed jammer,
the lemma has been shown.

\qedwhite\vspace{0.5cm}

Since classical arbitrarily varying channels can be regard as special cases of AVCQCs,
the following Corollary
\ref{mmiwip1}
(the classical capacity formula 
for classical AVC with informed  jammer) is a direct consequence
of Theorem \ref{lvvwatcv}.

\begin{corollary} \label{mmiwip1}
The 
$(V', V)$-correlation assisted  capacity of a 
classical AVC  ${\cal W}=\{\rho(x,s):x\in{\cal X}, s\in {\cal S}\}$ is equal to
 \begin{equation}\max_P  \min_{\bar{\bar{\rho}}(\cdot) \in \bar{\bar{W}}} 
I(P, \bar{\bar{\rho}}(\cdot)), \label{mmiwip}\end{equation} 
when $I(V', V)>0$. Here
$\bar{\bar{{\cal W}}}$
$:=\{\{\sum_s Q(s|x) \rho(x,s), x \in {\cal X}\}: \forall Q: {\cal X} \rightarrow {\cal S}\}$.
\end{corollary}

Instead of  deducing
Corollary
\ref{mmiwip1} from Theorem \ref{lvvwatcv},
we can also show this Corollary using a classical technique,
when we apply the classical
results of
Kiefer and Wolfowitz in 
\cite{KW62} directly to the real vector space of the channel
output, as 
mentioned in Section \ref{CotP}.
We  would like to skip a lengthy proof  
and leave the
extended proof of  Corollary \ref{mmiwip1} to the
readers as an exercise.
We  give just a sketch of the proof of Corollary
\ref{mmiwip1} by means of
the classical
results of
Kiefer and Wolfowitz as follows.

At first, similar to Lemma \ref{theorema},
we demonstrate that 
there exists  a hyperplane separating
 the classical channel outputs
 into two parts  for any correlated source $(V', V)$ with  $I(V', V)>0$.
 Then, we apply  the classical
results of
Kiefer and Wolfowitz in   \cite{KW62}
to construct a  binary point to
point classical channel with positive capacity,
as in Section \ref{APaaPftAAVClC}.
 Finally,
 to show 
the equality
of correlation assisted  capacity
and randomness assisted capacity, we
create 
common randomness  with a pre-code with negligible length,
as in the proof of Theorem \ref{lvvwatcv}.
 With this approach we can show the 
coding theorem for the arbitrarily varying  
classical channel.
\vspace{0.5cm}

\section{Applications} \label{lications}

Common randomness generation plays a fundamental role in various problems of cryptography and information
theory.
Here the channel users want to calculate a
shared random variable
using  an  AVCQC and
correlation as a resource.
This can be used, for instance, as a strong resource for a randomized protocol.
Obviously, the message transmission capacity of
any channel is upper bounded by its common randomness capacity.
Furthermore,  the common randomness  capacity of
any channel is upper bounded by its identification capacity.

Common randomness generating over a classical arbitrarily varying channel
using correlation
as a resource was first introduced in \cite{Ahl/Cs},
where it was assumed that the jammer had no
side information about the input codeword.
As an application of our results in Section \ref{mrap},
we now want to analyze the common randomness generation
using correlation
as resource for our  scenario in  Section \ref{mrap}, i.e.
we assume that the jammer knows the  input codeword.
We assume that the sender and the receiver
use a correlation assisted code to generate 
a pair of random variables $(J,L)$, both distributed on  a finite set ${\cal J}$,
so that the probability that
$J \not= L$
can be kept arbitrarily small.\vspace{0.2cm}

As a second  application of our results, 
we analyze the capacity formulas of this work and some of
our previous works to determine whether
they are Turing computable.

\subsection{Correlation Assisted 
Common Randomness Generation Capacity 
with an Informed Jammer  }
\label{CACRGCwaIJ}
In this Section, we consider how
much common randomness  an AVCQC with
an informed jammer can generate, using correlation as resource.
 We consider the two scenarios:
In one scenario 
the  channel users receive one signal from the correlation
for every channel use, while in the other scenario
 only a small number of signals from the correlation for mutable
channel uses are available  (cf. the discussion in Section 
\ref{CACF}).

\subsubsection{Capacity Definition and Communication Scenario}



\begin{definition}
Let ${\cal W}$ $=\{\{\rho(x,s), x \in {\cal X}\}, s \in {\cal S}\}$ be an  AVCQC
and   $(V', V)$ with alphabets $( {\cal V}',  {\cal V})$ be an arbitrary correlated source. 
 A non-negative number $R$ is an achievable   $(V', V)$-correlation 
 assisted 
common randomness generation rate   
with an informed jammer   
  for ${\cal W}$,
		if for every
$\epsilon>0$, $\delta>0$,  and sufficiently large $n$ there
exists a  random variable $J$ distributed  on  a finite set ${\cal J}$,
 a set of  encoders $\left\{u_{{v'}^{n}}:
{\cal J} \rightarrow {{\cal X}}^n:
{v'}^{n}\in{{\cal V}'}^{n}\right\}$ 
and a set of  measurements  $\left\{L_{{v}^{n}}:
\mathcal{S}({\cal H}^{\otimes n})\rightarrow {\cal J}
:{v}^{n}\in{{\cal V}}^{n}\right\}$
 such that $\frac{1}{n}H(J)>R-\delta$ and
\[\max_{ {\bf s}^n (\cdot) } Pr \left\{J  \not= 
 L\left( \rho\left(U_{{V'}^{n}},{\bf s}^{n}(U_{{V'}^{n}})\right), V^{n} \right)\right\} <\epsilon.\]
	The supremum on achievable	  $(V', V)$-correlation  assisted common randomness generation rate   
with an informed jammer of  ${\cal W}$
is called  the $(V', V)$-correlation assisted 
common randomness generation capacity 
with an informed jammer   
  for ${\cal W}$, denoted by $\tilde{C}({\cal W})$.
	\end{definition}
 
\begin{definition}
Let ${\cal W}$ $=\{\{\rho(x,s), x \in {\cal X}\}, s \in {\cal S}\}$ be an  AVCQC
and   $(V', V)$ with alphabets $( {\cal V}',  {\cal V})$ be an arbitrary correlated source. 
 A non-negative number $R$ is an achievable  
 $((V', V),(l_n)_{n\in\mathbb{N}})$-correlation assisted 
common randomness generation rate   
with an informed jammer   
  for ${\cal W}$,
		if for every
$\epsilon>0$, $\delta>0$,  and sufficiently large $n$ there
exists an   encoder  $\left\{u_{{v'}^{l_n}}:
{\cal J} \rightarrow {{\cal X}}^n:
{v'}^{l_n}\in{{\cal V}'}^{l_n}\right\}$,
and a set of   measurements   $\left\{L_{{v}^{l_n}}:
\mathcal{S}({\cal H}^{\otimes n})\rightarrow P({\cal J}) 
:{v}^{l_n}\in{{\cal V}}^{l_n}\right\}$,
 such that $\frac{1}{n}H(J)>R-\delta$ and
\[
\max_{ {\bf s}^n (\cdot) } Pr \Bigl\{J
\not= 
 L\left( \rho\left(U_{{V'}^{l_n}},{\bf s}^{n}(U_{{V'}^{l_n}})\right), V^{l_n} \right)\Bigr\} <\epsilon
\text{.}\]
		The supremum on achievable	on  $((V', V),(l_n)_{n\in\mathbb{N}})$-correlation 
assisted common randomness generation rate   
with an informed jammer   
  for ${\cal W}$ is called  the 
$((V', V),(l_n)_{n\in\mathbb{N}})$-correlation assisted 
common randomness generation capacity 
with an informed jammer   
  for ${\cal W}$, denoted by 
$\tilde{C}({\cal W}, (l_n)_{n\in\mathbb{N}})$.
	
 \end{definition}

\subsubsection{Capacity Results}

\begin{corollary}\label{vvinlirhlcrb}
Let $(V', V)$ with alphabets $({\cal V}', {\cal V})$  be an arbitrary correlated source 
and ${\cal W}=\{\rho(x,s):x\in{\cal X}, s\in {\cal S}\}$ be an AVCQC. 
   There is a positive $r''$ such that
for any  sequence of natural numbers $(l_n)_{n\in\mathbb{N}}$ such that $r''<$ $\liminf_{n\rightarrow\infty} \frac{l_n}{\log n}$ 
$\leq\limsup_{n\rightarrow\infty} \frac{l_n}{\log n}$ $< \infty$,
when $I(V', V)>0$ holds, then
 \begin{equation}\tilde{C}({\cal W}, (l_n)_{n\in\mathbb{N}})\geq \max_P  \min_{\bar{\bar{\rho}}(\cdot) \in \bar{\bar{W}}} 
\liminf_{n\rightarrow\infty} \frac{n-l_n}{n}\chi(P, \bar{\bar{\rho}}(\cdot))+\liminf_{n\rightarrow\infty} \frac{l_n}{n} r''.\end{equation}
\end{corollary}\vspace{0.2cm}

{\it Proof:} 
 We define $r'':= \frac{3}{r}$,
where $r$ is defined as in Lemma \ref{lemmaKW}.
We fix a $P$ and define \[J_n:=\lfloor 2^{ n\min_{\bar{\bar{\rho}}(\cdot) \in \bar{\bar{W}}} 
\chi(P, \bar{\bar{\rho}}(\cdot))-\delta} \rfloor\] for an arbitrary positive $\delta$.
Now the sender chooses a 
random variable uniformly distributed on $\{1,\cdots J_n\}$.
Let $(l_n)_{n\in\mathbb{N}}$ be a sequence such that
$\frac{\nu(n)}{\log n}$ $=r''$ $<\liminf_{n\rightarrow\infty} \frac{l_n}{\log n}$
$\leq\limsup_{n\rightarrow\infty} \frac{l_n}{\log n}$ $< \infty$.
   By Corollary \ref{awcvvlmrc} 
he can send the output to the receiver 
using an   $(n, l_n, J_n)$  code.
When the receiver fails to decode the output, he randomly chooses one element in
$\{1,\cdots J_n\}$. By Corollary \ref{awcvvlmrc}, when $I(V', V)>0$, the probability of failure with an informed jammer
can be
kept arbitrarily small when $n$ is sufficiently  large.

We define $L_{l_n}:=\lfloor 2^{l_n r} \rfloor$.
Now by classical common randomness generation technique in \cite{Ahl/Cs} 
we can define a set of  $L_{l_n}$
deterministic codes $\{{\cal C}^{\hat{W}}_{l}: l\in \{1,\cdots, L_{l_n}\}\}$ for $\hat{W}$, which is
the classical channel we defined in Lemma \ref{lemmaKW},
such that 
$L_{l_n}$ messages can be sent, and furthermore,
for ever random variable $K_s$ distributed on 
$\{{\cal C}^{\hat{W}}_{l}: l\in \{1,\cdots, L_{l_n}\}\}$,
the receiver can generate  random variable 
 $K_r$ distributed on  $\{1,\cdots, L_{l_n}\}$,  such that
when $l_n$ is sufficiently large, $Pr\{K_s\not= K_r\}<\epsilon$ for
any positive $\epsilon$.

Let $\mathcal{K}$ be the set in
the proof of Theorem \ref{lvvwatcv}, on which the common randomness takes value. 
$\{{\cal C}^{\hat{W}}_{l}: l\in \{1,\cdots, L_{l_n}\}\}$ defines a 
$(V', V)$-correlation assisted   code
\[\Bigl\{\Bigl(\left(u_{{v'}^{\nu(n)}}(k)\right)_{k=1,\cdots,\left|\mathcal{K}\right|},   \{D_{k}^{\nu(n)}:
k=1,\cdots,\left|\mathcal{K}\right|\}\Bigr)_{l}: l\in \{1,\cdots, L_{l_n}\}\Bigr\}\] with deterministic encoder $u_{{v'}^{\nu(n)}}(k) \in\{g_0^m,g_1^m\}^{\nu(n)}$
for ${\cal W}$ of
length $\nu(n)$ such that when $l_n$ is sufficiently large,
$L_{l_n}$ messages can be sent, 
and the receiver can generate a random variable  
 $K_r$  on  $\{1,\cdots, L_{l_n}\}$, with
 $Pr\{K_s\not= K_r\}<\epsilon$ for
any positive $\epsilon$.

We choose \[{\cal  J}:= \{1,\cdots J_n\} \times \{1,\cdots, L_n\}.\]
Now we can contract an $(n-l_n, l_n, J_n)$ 
two-part code as in the proof for Theorem 
\ref{lvvwatcv}, where the second part is a common randomness 
assisted code sending the random output of a 
variable uniformly distributed on $\{1,\cdots J_n\}$,
and the first part is used both to 
send $L_{l_n}$ messages 
(which are used as common randomness for the second part),
and to generate  random variables  $K_s$ and
 $K_r$, both uniformly distributed on  $\{1,\cdots, L_{l_n}\}$, with
 $Pr\{K_s\not= K_r\}<\epsilon$ for
any positive $\epsilon$. 
\qedwhite\vspace{0.3cm}

\begin{corollary}\label{vvinlirhlcrbn}
Let $(V', V)$ with alphabets $({\cal V}', {\cal V})$ be an arbitrary correlated source, 
and ${\cal W}=\{\rho(x,s):x\in{\cal X}, s\in {\cal S}\}$ be an AVCQC.

\begin{enumerate}
\item \label{itemb1}
 When 
\[I(V', V) \leq \max_P  \min_{\bar{\bar{\rho}}(\cdot) \in \bar{\bar{W}}} 
\chi(P, \bar{\bar{\rho}}(\cdot))\] 
holds, then
 \begin{equation}\tilde{C}({\cal W})= \max_P  \min_{\bar{\bar{\rho}}(\cdot) \in \bar{\bar{W}}} 
\chi(P, \bar{\bar{\rho}}(\cdot))+ I(V', V).\end{equation}
\item \label{itemb2} When 
\[I(V', V) > \max_P  \min_{\bar{\bar{\rho}}(\cdot) \in \bar{\bar{W}}} 
\chi(P, \bar{\bar{\rho}}(\cdot))\] 
holds, then
 \begin{equation}\tilde{C}({\cal W})=\sup_{U\rightarrow V'\rightarrow V} \Bigl\{I(U,V'): I(U;V')-I(U;V) \leq \max_P \min_{\bar{\bar{\rho}} (\cdot) \in \bar{\bar{\cal W}}}\chi(P, \bar{\bar{\rho}}(\cdot))  \Bigr\}.
\end{equation}\end{enumerate}
\end{corollary}\vspace{0.2cm}

{\it Proof:} 
 Our proof is based on the approach in  \cite{Ahl/Cs}
for correlation assisted
common randomness of classical channels.

At first we assume that 
$I(V', V) \leq \max_P  \min_{\bar{\bar{\rho}}(\cdot) \in \bar{\bar{W}}} 
\chi(P, \bar{\bar{\rho}}(\cdot))$ holds.

We fix a $P$ and define \[J_n':=\lfloor 2^{ n(\min_{\bar{\bar{\rho}}(\cdot) \in \bar{\bar{W}}} 
\chi(P, \bar{\bar{\rho}}(\cdot))+ I(V', V))-\delta } \rfloor\] for an arbitrary positive $\delta$. 
 Our idea is to have the transmitters share a random variable uniformly  distributed on $\{1,\cdots J_n'\}$
by means of  ${\cal W}$.

Let $\mathcal{K}$ be the set in the proof of Theorem \ref{lvvwatcv}
on which the common randomness takes value.
We choose a $|{\cal V}|$ dimensional Hilbert Space ${\cal H}^{|{\cal V}|}$ and 
a set of pairwise orthogonal pure quantum states
$\{\sigma_{v}:v\in V'\} \in \mathcal{S}( {\cal H}^{|{\cal V}|})$.
 We further define a map $g: {\cal V} \rightarrow \{\sigma_{v}:v\}$
by $g(v)= \sigma_{v}$. We have $\chi(V', g(\cdot)) = I(V', V)$.
We now generate 
\[ \left|\mathcal{K}\right|\left|{\cal V}'\right|\lfloor 2^{ n(\min_{\bar{\bar{\rho}}(\cdot) \in \bar{\bar{W}}} 
\chi(P, \bar{\bar{\rho}}(\cdot))+  I(V', V))-\delta} \rfloor\]
random variable 
\[\{U_{k,v'}(j): j\in\{1,\dots J_n'\}, k\in k\in \mathcal{K}, v' \in {\cal V}'	\}, \]
u in ${\cal X}$.
Since 
\[\chi(P\times V'; \bar{\bar{\rho}}(\cdot)\otimes g(\cdot)) \geq   \chi(P; \bar{\bar{\rho}}(\cdot)) + I(V', V)\]
by \cite{Bo/Ca/Ca} when $n$ is sufficiently large, then with
 a positive probability 
according to the joint distribution of 
$V'$, $V$, and the uniform distribution on $\mathcal{K}$,
there is  a map $g: {\cal V} \rightarrow \{\sigma_{v}:v\}$, a realization
$\{u_{k,v'}(j): j, k, v'	\} $ of 
$\{U_{k,v'}(j): j, k, v'	\} $, and a family of
decoding sets 
\[\Bigl\{\{{\cal D}(j,k,g(v)), j \in \{1,\dots J_n'\}\}, k \in {\cal K}, v \in {\cal V}\Bigr\}\]
 such that for any positive $\epsilon$ and sufficiently large $n$
\begin{equation}\min_{{\bf s}}\frac{1}{J_n'} \sum_{j=1}^{J_n'}
\mathbb{E}tr[\rho^{\otimes n}({\bf u}(j,K,V'), {\bf s} ({\bf u}(j,K,V'))){\cal D}(j,K,V)] > \epsilon\label{mbs1jkv}\end{equation}
according to the joint distribution of
$V'$, $V$, and the uniform distribution on $\mathcal{K}$
with an informed jammer.

Now we can contract an $(n, J_n')$ 
two-part code.
By Corollary \ref{vvinlirhlcrb}
we can
define a 
deterministic code of negligible length
such that 
$\left|\mathcal{K}\right|$ messages can be sent.
The first part of the codeword 
sends $\left|\mathcal{K}\right|$ messages 
 as common randomness for the second part.
The second part is the  randomness 
assisted code 
defined in (\ref{mbs1jkv}),
sending the random output of a 
variable distributed on $\{1,\cdots J_n'\}$.
This shows the direct part for this case.

For the converse, we 
suppose that after the message transmission
the sender and the receiver share a 
random variable $M$ which is independent of
$V'$ and $V$. We now
consider
the Markov chain
$M\rightarrow PV' \rightarrow \rho(\cdot)V \rightarrow M$.
By the data processing inequality and the capacity formula for
 $\rho$, with an informed jammer in \cite{Bo/Ca/Ca}, we have
\[H(M) \leq \chi(P, \bar{\bar{\rho}}(\cdot))+ I(V', V).\]
\vspace{0.2cm}

Now  we assume that 
$I(V', V) > \max_P  \min_{\bar{\bar{\rho}}(\cdot) \in \bar{\bar{W}}} 
\chi(P, \bar{\bar{\rho}}(\cdot))$ holds. 

Similar to above,
we fix a $P$ and define \[J_n'':=\lfloor 2^{ n(\min_{\bar{\bar{\rho}}(\cdot) \in \bar{\bar{W}}} 
\chi(P, \bar{\bar{\rho}}(\cdot))+ I(U; V))-\delta } \rfloor\] for an arbitrary positive $\delta$
and  an
$U\rightarrow V'\rightarrow V$ such that $I(U;V')-I(U;V) \leq \chi(P, \bar{\bar{\rho}}(\cdot))$.

We now generate 
\[ \left|\mathcal{K}\right|\left|{\cal V}'\right|\lfloor 2^{ n(\min_{\bar{\bar{\rho}}(\cdot) \in \bar{\bar{W}}} 
\chi(P, \bar{\bar{\rho}}(\cdot))+  I(U, V))-\delta} \rfloor\]
random variable 
\[\{U_{k,v'}(j): j\in\{1,\dots J_n''\}, k\in k\in \mathcal{K}, v' \in {\cal V}'	\}.\]
Similar to above, we want to have the transmitters share a random variable uniformly  distributed on $\{1,\cdots J_n''\}$
by means of  ${\cal W}$.

When $U\rightarrow V'\rightarrow V$ holds, then it also holds that
$U\times P \rightarrow V'\times P \rightarrow g(V)\otimes \rho(P)$.
Since 
\[\chi(P\times U; \bar{\bar{\rho}}(\cdot)\otimes g(\cdot)) \geq   \chi(P; \bar{\bar{\rho}}(\cdot)) + I(U, V)\]
when $n$ is sufficiently large, then with
 a positive probability 
according to the joint distribution of 
$V'$, $V$, and the uniform distribution on $\mathcal{K}$,
there is  a map $g: {\cal V} \rightarrow \{\sigma_{v}:v\}$, a realization
$\{u_{k,v'}(j): j, k, v'	\} $ of 
$\{U_{k,v'}(j): j, k, v'	\} $, and a family of
decoding sets 
\[\Bigl\{\{{\cal D}(j,k,g(v)), j \in \{1,\dots J_n''\}\}, k \in {\cal K}, v \in {\cal V}\Bigr\}\]
 such that for any positive $\epsilon$ and sufficiently large $n$
\[\min_{{\bf s}}\frac{1}{J_n''} \sum_{j=1}^{J_n''}\mathbb{E}tr[\rho^{\otimes n}({\bf u}(j,K,V'), {\bf s} ({\bf u}(j,K,V'))){\cal D}(j,K,V)] > \epsilon\]
according to the joint distribution of
$V'$, $V$, and the uniform distribution on $\mathcal{K}$.

   Similar to above, we can contract an $(n, J_n'')$ 
two-part code, where
the first part is used to 
send $\left|\mathcal{K}\right|$ messages and
the second part is a  randomness 
assisted code 
sending the random output of a 
variable distributed on $\{1,\cdots J_n''\}$.

For the converse, we consider a $U$ with
$U \rightarrow V'\rightarrow V$.
Let $Y$ be the classical random outcome of the decoding measurement.
By the data processing inequality 
and the capacity formula for
 $\rho$ with an informed jammer in \cite{Bo/Ca/Ca}, we have
\[I(X;Y) \leq \chi(P, \bar{\bar{\rho}}(\cdot)).\]
We now apply the results for common randomness capacity 
via classical channel in \cite{Ahl/Cs} on the resulting classical 
arbitrarily varying channel with an informed jammer
$X\rightarrow Y$. We have
\begin{align*}&
\tilde{C}({\cal W})\\
&\leq \sup_{U\rightarrow V'\rightarrow V} \Bigl\{I(U,V'): I(U;V')-I(U;V) \leq I(X;Y)  \Bigr\}\\
&\leq  \sup_{U\rightarrow V'\rightarrow V} \Bigl\{I(U,V'): I(U;V')-I(U;V) \leq \max_P \min_{\bar{\bar{\rho}} 
(\cdot) \in \bar{\bar{\cal W}}}\chi(P, \bar{\bar{\rho}}(\cdot))  \Bigr\}	\text{ .}\end{align*}
 \qedwhite\vspace{0.5cm}

\subsection{Computability}

As an application, 
we now consider whether the  capacity formulas in Section \ref{mrap}
and Section \ref{CACRGCwaIJ} can be calculated algorithmically
in finite time, or whether they are computable.

\subsubsection{Capacity Definition and Communication Scenario}

\begin{definition}
A sequence of rational numbers $\{r_n: n\in\mathbb{N}\}$ is called a computable
sequence if there exist recursive functions $a$, $b$, and $s$ $: \mathbb{N}\rightarrow
\mathbb{N}$ such that for all $n\in\mathbb{N} $ we have
$b(n) \not= 0$  and
\[r_n=(-1)^{s(n)}\frac{a(n)}{b(n)} .\]

 \end{definition}

\begin{definition}
A function $f$ : $\mathbb{R}_c \rightarrow \mathbb{R}_c$ 
 is called Banach-Mazur computable if it maps any 
computable sequence
$\{r_n: n\in\mathbb{N}\}$ of real numbers into a computable sequence
$\{f(r_n): n\in\mathbb{N}\}$ of real numbers.
Here $\mathbb{R}_c$, the set of
computable
numbers,  is defined as the set of
 real numbers that are computable by Turing
machines.
\end{definition}

\begin{definition}
A rapidly converging Cauchy representation of
a computable real $x$ is a sequence of real numbers
 $\{x_n: n\in\mathbb{N}\}$ that
converges to $x$ when $n\rightarrow\infty$ rapidly, i.e. for every $i$ and $j\geq i$
 it holds $|x_j-x_i|<2^{-i}$.
\end{definition}

\begin{definition}
A function $f$ : $\mathbb{R}_c \rightarrow \mathbb{R}_c$ is called Borel computable
if there is an algorithm that transforms each given
rapidly converging Cauchy representation of a computable
real $x$ into a corresponding representation for $f(x)$.
\end{definition}

Notice that Borel computability implies Banach-Mazur computability.

 \begin{definition}
An AVCQC ${\cal W}$
 with  input alphabet ${\cal X}$ and output space ${\cal H}$
 is  computable
 if it maps  every letter in ${\cal X}$
quantum state in $\mathcal{S}({\cal H})$
with 
computable coefficient in $\mathbb{C}$.
\end{definition}

\begin{definition} Assume we have two
 random variables
$(V',V)$ and $(\dot{V}',\dot{V})$ both taking values in a finite set ${\cal V}' \times  {\cal V}$ 
with joint distributions  $P_{V',V}$ and $P_{\dot{V}',\dot{V}}$,  respectively.
We define their distance by
\[\|P_{V',V}-P_{\dot{V}',\dot{V}}\|_{1}=\sum_{v'\in{\cal V}'}\sum_{v\in{\cal V}}
|P_{V',V}(v',v)-P_{\dot{V}',\dot{V}}(v',v)|.\]

Assume we have  two AVCQCs,
${\cal W}$ $=\{\{\rho(x,s), x \in {\cal X}\}, s \in {\cal S}\}$ and
$\dot{{\cal W}}$ $=\{\{\dot{\rho}(x,s), x \in {\cal X}\}, s \in {\cal S}\}$, 
with  correlated sources
$(V',V)$ and $(\dot{V}',\dot{V})$  with alphabets $( {\cal V}',  {\cal V})$,  respectively.
We define a distance of
$\Bigl({\cal W},(V',V)\Bigr)$
and $\Bigl(\dot{{\cal W}},(\dot{V}',\dot{V})\Bigr)$
by
\[ d\biggl(\Bigl({\cal W},(V',V)\Bigr), \Bigl(\dot{{\cal W}},(\dot{V}',\dot{V})\Bigr)\biggr)
:=\|{\cal W}-\dot{{\cal W}}\|_{\lozenge}+\|P_{V',V}-P_{\dot{V}',\dot{V}}\|_{1}.\]
Here \[ \|W \|_{\lozenge}:=\sup_{n\in \mathbb{N}}\max_{a\in S(\mathbb{C}^n
\otimes H'), \|a\|_1=1}\| (\mathrm{id}_n \otimes W)(a)\|_1\text{ .}\]
\label{t2awoniah}
\end{definition}

\subsubsection{Capacity Results}

 As an application of our capacity results,
in this section we want to
analyze whether the capacity formulas
determined in  Section \ref{mrap}
and in Section \ref{CACRGCwaIJ} 
are computable functions of the channel 
parameters on a Turing machine or not.

Our computability analysis is novel because,
according to our knowledge, most of the capacity results,
e.g. correlation assisted 
common randomness generation capacity 
with an informed jammer, do not yet exist for
classical channels. Furthermore,
the deterministic
capacity formula using maximal error criterion of
 arbitrarily varying channels
with an informed jammer is still unknown,  since it
contains  Shannon's
zero-error capacity, which is still an open problem 
 as a special case (cf. \cite{Bo/Ca/Ca}).
Corollary \ref{ctivvcwaa2} shows that even if we
do not know the capacity formula, we can predict
that no  computable formula can ever be found
(cf. Remark \ref{Nttcfwi0}).

We would also like to 
point out
that, since classical arbitrarily varying channels
can be regard as special cases of AVCQCs, 
the computability of a
classical-quantum capacity formula 
implies the computability of 
its corresponding classical capacity formula,
while the non-computability of a
classical capacity formula 
implies the non-computability of 
its corresponding classical-quantum capacity formula.
Thus, for any  classical-quantum capacity formula, 
which is proved to be computable in this section,
it is clear that the  corresponding classical capacity formula 
is automatically computable also.
On the other hand,
to show that a classical-quantum capacity formula is not 
computable,
it is sufficient to show that  the  corresponding classical capacity formula
is not 
computable.


\begin{corollary}\label{ctivvcwaa}
Let $(V', V)$ with alphabets $({\cal V}', {\cal V})$ be an arbitrarily correlated source, 
and ${\cal W}=\{\rho(x,s):x\in{\cal X}, s\in {\cal S}\}$ be an AVCQC. 
When $I(V', V)>0$ holds, then
$C( {\cal W};corr(V', V))$
is Turing computable.
\end{corollary}

{\it Proof:} 
By Theorem \ref{lvvwatcv}, when $I(V', V)>0$ holds, then
we have
\[C( {\cal W};corr(V', V))=\max_P  \min_{\bar{\bar{\rho}}(\cdot) \in \bar{\bar{W}}} 
\chi(P, \bar{\bar{\rho}}(\cdot)).\]
$\chi(P, \bar{\bar{\rho}}(\cdot))$ is a 
 Turing computable continuous function, and 
the minimum of a   computable continuous function
on a  computable set is also a   computable continuous function.
Thus $C( {\cal W};corr(V', V))$ is Turing computable.\qedwhite\vspace{0.4cm}

%

\begin{corollary}\label{ctivvcwaa2}
Let $(V', V)$ with alphabets $({\cal V}', {\cal V})$ be an arbitrarily correlated source, 
and ${\cal W}=\{\rho(x,s):x\in{\cal X}, s\in {\cal S}\}$ be an AVCQC. 
When $I(V', V)=0$ holds, then
$C( {\cal W};corr(V', V))$ using maximal error criterion on the message set
is, in general, not Turing computable.
\end{corollary}

{\it Proof:} 
We show Corollary \ref{ctivvcwaa2} by a counterexample for a classical AVC, since
classical  arbitrarily varying channels can be
regarded as special cases of AVCQCs. 

Let ${\cal X}' = {\cal Y}' = \{a, 0, 1, 2\}$ and ${\cal S} = \{s_0, s_1\}$.
We define a  classical  arbitrarily varying channel ${\cal W}' $
such that $W'(a|a, s_0) = W'(a|a, s_1) = 1$, $W'(y|x; s_i) = 1$ if
$y = x+i~(mod~3)$  for $x, y \in \{0, 1, 2\}$. That is, the transmission
matrices in ${\cal W}'  $ are
\[\left ( \begin{array} {rrrr}
1&0&0&0\allowdisplaybreaks\\
0&1&0&0\allowdisplaybreaks\\
0&0&1&0\allowdisplaybreaks\\
0&0&0&1\allowdisplaybreaks\\ \end{array}\right )\text{ ,}
~~\left ( \begin{array} {rrrr}
1&0&0&0\allowdisplaybreaks\\
0&0&1&0\allowdisplaybreaks\\
0&0&0&1\allowdisplaybreaks\\
0&1&0&0\allowdisplaybreaks\\ \end{array}\right )\text{ .}\]
In our previous work  \cite{Bo/Ca/Ca}, we have shown that
 the randomness assisted capacity of
${\cal W}'$ with an informed jammer  using maximal error criterion is 
equal to $\log \frac{5}{2}$.
By  Theorem \ref{lvvwatcv},
when we have a random variable pair $B$
and $B'$ with $I(B, B')>0$,   then
$C( {\cal W}';corr(B, B'))$ using maximal error criterion is 
also equal to $\log \frac{5}{2}$.
Furthermore, in \cite{Bo/Ca/Ca}  we have shown that 
the deterministic capacity (i.e. with no resource) of
${\cal W}'$ with an informed jammer 
using
maximum error probability is  equal to  $2$.

Let $r\in [0,1]$.
Let ${\cal X}'' = {\cal Y}'' = \{ 0, 1\}$.
We define a  classical  channel ${\cal W}_r'' $
such that $W_r''(0|0) = W_r''(1|1) = 1$
and $W_r''(0|1) = W_r''(1|0) = 1-r$.
It is easy to see that the capacity of
 ${\cal W}_r''$ is zero if and only if $r=0$.

We now define  ${\cal X} = {\cal Y} := {\cal X}' \times {\cal X}''$,
  and ${\cal S} = \{s_0, s_1\}$.
We define a  classical  arbitrarily varying channel ${\cal W}_r := {\cal W }_r''\otimes{\cal W}'$,
that is $W_r((y,y')|(x,x'); s_i) = W_r''(y|x)\cdot W'(y'|x'; s_i)$.

When $r>0$ holds, the capacity of
 ${\cal W}_r''$ is positive, thus there 
is a distribution $B'$ on ${\cal X}'' $
such that $I(B', {\cal W}_r''(B'))>0$.
It follows directly that there is  a
 distribution $B$ on ${\cal X}$
such that $I(B,W_r(B))>0$.
Similar to the proof of  Theorem \ref{lvvwatcv},
we may use a negligible amount of bits to create
common randomness. Notice that since we only demand maximal error criterion on the message set,
for the common randomness we may use the average error criterion. This means that for any random variable pair $V'$
and $V$, even when $I(V', V)=0$ holds, 
$C( {\cal W}_r;corr(V', V))$ using maximal error criterion is always larger or equal to
$\log \frac{5}{2}$. 

However, when $r=0$, the capacity of
 $\cal V $ is zero. In this case we have 
$C( {\cal W}_0;corr(V', V))$
$ = C( {\cal W}';corr(V', V))$.
When $I(V', V)=0$ holds, we have no resource, then
$C( {\cal W}_0;corr(V', V))$ using maximal error criterion
is   equal to  $2$.

Thus $C( {\cal W};corr(V', V))$ using maximal error criterion
is not a continuous function of the channel parameters  when
$I(V', V)=0$. 
Since a Banach-Mazur computable function
is continuous on the computable sets,
$C( {\cal W};corr(V', V))$ using maximal error criterion is 
not Banach-Mazur computable 
and thus is not
Turing computable 
(cf. \cite{Bo/Sch/Ba/Po}). 
\qedwhite\vspace{0.2cm}

\begin{remark}
Usually the capacity formulas of 
 arbitrarily varying channels are
discontinuous in zero, i.e.
on the discontinuity points,
positive capacity sinks rapidly to zero.
Corollary \ref{ctivvcwaa2}
is a rare example where the discontinuity point
is positive. Here, when $r\rightarrow 0$, 
the capacity has a fluctuation, but remains positive.
\end{remark}

\begin{remark}\label{Nttcfwi0}
Notice that the capacity formula  when $I(V', V)=0$, i.e. the deterministic
capacity formula   using maximal error criterion of
 arbitrarily varying channels
with an informed jammer is still an open problem, even for classical channels, since it
contains
the zero-error capacity of related discrete memoryless
channels as a special case (cf. \cite{Bo/Ca/Ca}).
Corollary \ref{ctivvcwaa2} shows that we can make a statement about the
computability, even when we do not have a formula for this quantity.
\end{remark}





\begin{theorem} If the condition \ref{itemb1} of
Corollary \ref{vvinlirhlcrbn}
is satisfied, then
$\tilde{C}({\cal W})$ is Banach-Mazur computable 
and
Turing computable. 
\end{theorem}

{\it Proof:}
We define
 \begin{equation}\Phi({\cal W}, P_{(V', V)}) := \max_P  \min_{\bar{\bar{\rho}}(\cdot) \in \bar{\bar{W}}} 
\chi(P, \bar{\bar{\rho}}(\cdot))+ I(V', V). \end{equation}
If \ref{itemb1}
is satisfied, then by Corollary \ref{vvinlirhlcrbn} we have

\[\tilde{C}({\cal W})=\Phi({\cal W}, P_{(V', V)}) .\]

The two expressions 
$\chi(P, \bar{\bar{\rho}}(\cdot))$ and
$I(V', V)$ are both Turing computable continuous functions. 
Since the minimum of a   computable continuous function
on a  computable set is also a   computable continuous function,
$\max_P  \min_{\bar{\bar{\rho}}(\cdot) \in \bar{\bar{W}}} 
\chi(P, \bar{\bar{\rho}}(\cdot))$ is Turing computable.
$\Phi({\cal W}, P_{(V', V)})$ is the sum of two 
computable functions, and thus also a computable function.
\qedwhite\vspace{0.4cm}

\begin{theorem}
 $\tilde{C}({\cal W}, (l_n)_{n\in\mathbb{N}})$ is in general
not Banach-Mazur computable 
and thus not
Turing computable.\label{tacsvvwivvso}
\end{theorem}\vspace{0.3cm}





{\it Proof:}

Let ${\cal V}'={\cal V}=\{1,2\}$. For $n\in\mathbb{N}$ 
we consider a correlated source 
$(V_{n}', V_{n})$ with alphabets $( {\cal V}',  {\cal V})$ 
and joint distributions  $P_{V_{n}', V_{n}}$
\[\left( \begin{array} {cc}\frac{1}{2}-\frac{1}{2^n}&\frac{1}{2^n}\\
\frac{1}{2^n}&\frac{1}{2}-\frac{1}{2^n}
 \end{array}\right)\text{ .}\]
We have $I(V_{n}', V_{n})>0$ for all $n>1$.
Let $(V', V)$ be the  correlated source 
on $( {\cal V}',  {\cal V})$ with joint distributions  $P_{V', V}$
\[\left( \begin{array} {rr}\frac{1}{2}&0\\
0&\frac{1}{2}
 \end{array}\right)\text{ .}\]
 We have \[\lim_{n\rightarrow\infty}\| (V_{n}', V_{n})-(V', V)\|_{1}=0 .\]

We consider an
 AVCQC ${\cal W}=\{\rho(x,s):x\in{\cal X}, s\in {\cal S}\}$ 
 such that for all 
$s\in {\cal S}$ and all $x\in{\cal X}$ we have
\[\rho(x,s)= \delta\]
for a  fixed quantum state	$\delta\in\mathcal{S}({\cal H})$. 
It holds
\begin{align*}&
\lim_{n\rightarrow\infty} d\biggl(\Bigl({\cal W},(V',V)\Bigr), \Bigl({\cal W},(V_{n}', V_{n})\Bigr)\biggr)\\
&=\lim_{n\rightarrow\infty}\| (V_{n}', V_{n})-(V', V)\|_{1}=0 .\end{align*}
The correlation assisted message transmission capacity is always zero,
even when the jammer has no side information about
the input. Furthermore, any classical channels which
arise from  ${\cal W}$ also have zero capacity.
Thus the $(V_{n}', V_{n})$-correlation assisted 
common randomness capacity of  ${\cal W}$ with an informed jammer   is
equal to the $(V_{n}', V_{n})$-correlation assisted 
common randomness capacity of a  useless classical arbitrary varying channel
without jamming attack.
Here, useless classical channel means a
classical channel 
with zero message transmission capacity.
By \cite{Wit},
the $(V_{n}', V_{n})$-correlation assisted 
common randomness capacity of any useless classical arbitrary varying channel
without jamming attack is equal to zero, 
thus 
 $\tilde{C}({\cal W}, (V_{n}', V_{n}))=0$ for   $n>1$.
Furthermore, it is not hard to see that
$\tilde{C}({\cal W}, (V', V))=1$.
Thus $\tilde{C}({\cal W}, (V', V))$ 
and $\tilde{C}({\cal W}, (V_{n}', V_{n}))$ for all $n>1$
lie in the set of computable numbers.
When we let $n$ 
tend to infinity, 
the correlation assisted 
common randomness capacity is discontinuous 
on $({\cal W}, (V', V))$.
Since a Banach-Mazur computable function
is continuous on the computable sets,
the correlation assisted 
common randomness capacity is 
not Banach-Mazur computable 
and thus not
Turing computable 
(cf. \cite{Bo/Sch/Ba/Po}).




\qedwhite\vspace{0.4cm}

\begin{corollary}
The $((V', V),(l_n)_{n\in\mathbb{N}})$-correlation assisted 
common randomness generation capacity  of an  AVCQC ${\cal W}$
with an uninformed jammer, i.e. when the jammer has no information
about  the input code word,   is in general,
not Turing computable.

\end{corollary}\vspace{0.3cm}

{\it Proof:}
In the no-computable
example which we give in
the proof of Corollary
\ref{tacsvvwivvso},
the correlation assisted message transmission capacity is always zero,
even when the jammer has no side information about
the input. 
Thus the $(V_{n}', V_{n})$-correlation assisted 
common randomness capacity of  ${\cal W}$ with an uninformed jammer   is
equal to the $(V_{n}', V_{n})$-correlation assisted 
common randomness  generation capacity of an arbitrary useless classical channel
without jamming attack.
By the proof of Corollary
\ref{tacsvvwivvso}, the $(V_{n}', V_{n})$-correlation assisted 
common randomness  generation capacity is not Turing computable
even when the jammer has no side information about
the input.


\qedwhite\vspace{0.4cm}

\begin{corollary}
The $((V', V),(l_n)_{n\in\mathbb{N}})$-correlation assisted 
common randomness  generation capacity  of a classical-quantum channel $\rho$
with no jammer   
is, in general, not Turing computable.

\end{corollary}\vspace{0.3cm}

{\it Proof:}
In the no-computable
example which we give  in
the proof of Corollary
\ref{tacsvvwivvso},
we consider that there is no
jamming attack.
Thus  $(V_{n}', V_{n})$-correlation assisted 
randomness generation
over classical-quantum channels
with no jammer contains 
this example 
as a special case.

\qedwhite\vspace{0.4cm}

\section*{Acknowledgment}
The work of H. Boche and N. Cai was supported by the
Gottfried Wilhelm Leibniz program of the German Research Foundation
(DFG) via Grant BO 1734/20-1 and  partly supported by the German Research Foundation
(DFG) under Germany's Excellence Strategy EXC-2111-390814868. The work of M. Cai was supported
by the  Bundesministerium f\"ur Bildung und Forschung (BMBF) via
Grant 16KIS0118K and
partly supported by the national research initiative on quantum
technologies of Bundesministerium f\"ur Bildung und Forschung (BMBF), 
within the project QuaDiQua 16KIS0948. 
The research direction on jamming and active attacks
on communication
systems was initiated by the German Research Foundation (DFG) under
Grants BO 1734/24-3 and BO 1734/25-1.\vspace{0.5cm}

\end{document}